\documentclass[aps,pra,showpacs,floatfix,superscriptaddress]{revtex4-1}
\usepackage{amsmath,amsfonts,amssymb,graphicx,float,footnote}
\usepackage[figtopcap]{subfigure}

\usepackage{times}

\begin{document}

\title{
  Dynamics of Non-classicality Measures in the Decohering Harmonic Oscillator
}
\author{Peter A. Rose}
\author{Andrew C. McClung}
\author{Tyler E. Keating}
\author{Adam T. C. Steege}
\affiliation{
  Department of Physics and Astronomy, Carleton College, 
  One North College Street, Northfield, MN 55057
}
\author{Eric S. Egge}
\affiliation{
  Department of Mathematics, Carleton College, 
  One North College Street, Northfield, MN 55057
}
\author{Arjendu K. Pattanayak}
\affiliation{
  Department of Physics and Astronomy, Carleton College, 
  One North College Street, Northfield, MN 55057
}
\date{
 \today 
}

\begin{abstract}

We show that eigenstates $|n\rangle$ of the harmonic oscillator coupled to a linear Markovian 
bath demonstrate a non-trivial behavior in the dynamics of measures of non-classicality. 
Specifically, as the system undergoes decoherence, a time-dependent peak in non-classicality 
as a function of $n$ emerges. We find this effect studying the dynamics of several 
non-classicality measures previously presented in the 
literature\cite{hillery87,marian02,dodonov03,kenfack04} which compare quantum states to 
the set of all classical states. In studying these measures we introduce a novel set 
of classical states for the purpose of calculations which improve upon the results obtainable using these measures. In addition, 
following in the footsteps of \cite{kenfack04}, we show that the negative volume of the 
Wigner function agrees well with all the non-classicality measures while being otherwise 
calculationally significantly more tractable. Finally, we explore the
dynamics of non-classicality of several other states. \end{abstract}

\maketitle

\section{Introduction}
The macroscopic world is constructed of fundamentally quantum objects, but the observation of 
quantum effects requires either very small or very cold systems. That is, quantum effects are most 
easily visible in systems sufficiently isolated from their environments to prevent decoherence. 
It is generally more difficult to adequately protect larger systems from decoherence; this difficulty 
gives rise to the classical behavior of the macroscopic scale \cite{zurek03}. 
Contemporary investigations of the transition from quantum to classical are motivated by
fundamental considerations as well as practical issues of control and engineering.
The transition is being investigated in nanomechanical systems 
\cite{blencowe04,brennecke08}, mesoscopic systems such as Josephson junction devices 
\cite{wal00}, as well as cavity--QED (quantum electrodynamics) systems \cite{mabuchi02}, among others.
This quantum-classical transition is controlled by 
internal system parameters and behaviors as well as (a) the relative size of $\hbar$ compared to
the characteristic action of the system, (b) the strength of the system-environment interaction, and
(c) the temperature and other characteristics of the environment, leading to a rich landscape of 
possible behaviors \cite{pattanayak03, mao10}.

The decohering harmonic oscillator has been the subject of many studies
over the years\cite{paavola11,janszky90,lu89,serafini04,brune08}. 
While these studies are highly relevant for the fundamental understanding 
of quantum mechanics itself, it is not just a theoretical exercise 
since there have been proposals to encode qubits and qudits 
(relevant in quantum information theory) in the harmonic oscillator 
using finite superpositions of eigenstates \cite{bartlett02}. In 
addition, experimental data now exists on the decay of these states 
where Brune et.\ al.\ successfully excited Fock states in a cavity-QED 
system, and monitored their decay using tomography \cite{brune08}.

In this paper we consider the quantum to classical transition in a decohering harmonic 
oscillator specifically studying the dynamics of measures of
``non-classicality" in this system. 
There have been several careful studies of non-classicality measures in static, closed systems, 
and other detailed studies of decoherence in open systems; however, to the best of our knowledge 
there has been only one previous study of the dynamics of measures of 
non-classicality \cite{paavola11} which looked at a restricted class of such measures on a 
restricted class of states. The study of these non-classicality measures in dynamical situations
(that is, in open decohering systems) allows for deeper insight into the
dynamics of decoherence while simultaneously providing a stronger test of
the meaning and validity of these non-classicality measures.
Our focus is on the harmonic oscillator eigenstates (the Fock or number
states) and we also consider finite superpositions of these states.  

We show below that indeed these measures of non-classicality have non-trivial 
dynamics for an open system. Specifically, as the system evolves, a time-dependent 
peak in ``non-classicality" as a function of eigenstate $n$ emerges. In studying
this behavior we also compare several measures of non-classicality and show that 
a somewhat intuitive measure (the negative volume of the Wigner function 
representation of the density matrix) works as well as or better than the other 
more formal measures.

We start our presentation of these results in Section II with a new
derivation of an analytical solution in the number basis of the behavior of the 
density matrix of a harmonic oscillator coupled to a linear Markovian bath. 
Although similar solutions have been previously derived, representing it in 
the number basis allows us to conveniently compute non-classicality as a 
function of time using several measures. Most of these measures are formulated 
in Hilbert-space, as opposed to characteristic or Wigner space.  Thus the 
convenience comes from the fact that our solution allows for extremely fast 
calculation of the individual elements of the density matrix for any time and 
any bath temperature. In Section III we discuss in some detail 
how non-classicality has been defined and quantified.  We first present 
different historical perspectives on the matter in order to gain intuition for
the concept, and then go on to present four formal definitions of non-classicality
from the literature which we believe provides a representative sample of the 
different techniques currently in use.  

Section IV is the heart
of our discussion where we present results, starting with the static
case; in the process of analyzing these, we are able to generalize
previous results in that we find a novel classical basis that improves upon 
previous calculations for the static case (and is later of value for the
dynamical case).  
We then present our results for decaying Fock states coupled to the 
vacuum, showing that during the transition from the quantum to the classical regime, 
a time dependent peak in non-classicality exists as a function of
eigenstate $n$. We discuss how these results align with our intuition
about classical-quantum correspondence. 
In addition, these results allow us to validate previous arguments that the 
negative volume of the Wigner function is a useful and meaningful measure of 
non-classicality, in that it agrees well with all the other non-classicality measures while being 
otherwise computationally significantly more tractable.  The possibility that the Wigner function
is a meaningful measure of non-classicality has current significance particularly because
suggestions of how to directly measure the Wigner function, put forward two decades ago,
are now being put into practice in many studies \cite{wilkens91,smithey93,singh10,mallet11}.
This suggests that non-classicality could be easily measured in 'real
time' without the need for detailed analysis.  Indeed, since several of the measures of 
non-classicality we study require computationally expensive searches --
that are fundamentally mathematically not guaranteed to have converged
-- we have relied on the negative volume of the Wigner function for our
analysis of non-zero temperature environments for number states and their finite 
superpositions. 
We close with a short conclusion and discussion of our findings in Section V.

\section{The Master Equation and Solution}

Consider a quantum harmonic oscillator linearly coupled to a thermal bath in the 
interaction picture.  Assuming that the bath is composed of a continuum of oscillators, 
that the back-action of the system on the bath is negligible, and modeling the interaction 
in the Markovian approximation, the time evolution of the density matrix $\rho$ is 
governed by the master equation \cite{gardiner10}
\begin{equation}
\label{master}
  \dot{\rho} = \frac{\gamma}{2}\left(
    NL[a^\dagger]\rho
    + (N+1)L[a]\rho
  \right).
\end{equation}
Here the dot represents the time derivative, the Lindblad superoperator is defined as $L[O]\rho 
\equiv2O\rho O^\dagger-O^\dagger O\rho-\rho O^\dagger O$, $a^\dagger$ and $a$ are the raising 
and lowering operators for the harmonic oscillator, respectively, $\gamma$ represents the degree 
of  coupling of the oscillator to its environment, and $N$ corresponds to the mean number of 
thermal photons in the bath.

We study this problem in the number basis, where the generally complex numbers  
$\langle n|\rho|m\rangle = C_{n,m}$ are the elements of $\rho$. In this representation 
equation (\ref{master}) becomes
\begin{widetext}
\begin{equation}
\label{numberbasis}
\frac{1}{\gamma}\dot{C}_{n,m} = \sqrt{n+1}\sqrt{m+1}(N + 1)C_{n+1,m+1} -
(nN +mN + N + \frac{n+m}{2}) C_{n,m} +\sqrt{n}\sqrt{m}N C_{n-1,m-1},
\end{equation}
\end{widetext}
making it clear that the diagonals decouple; that is, $C_{n,m}$ does not depend on $C_{n+i,l+i}$ or 
$C_{j+i,m+i}$, where $i,j,l,m$, and $n$ are integers, $l\neq m$ and $j\neq n$ \cite{gardiner10}.
This fact allows us to make the substitution $m=n+k, k\geq0$ into Eq.~(\ref{numberbasis}), yielding
\begin{equation}
\nonumber
\frac{1}{\gamma}\dot{C}_{n,n+k} = \sqrt{n+1}\sqrt{n+k+1}(N + 1)C_{n+1,n+k+1}
\end{equation}
\begin{equation}
\label{numberbasisk}
- ((2n+k+1)N + \frac{2n+k}{2}) C_{n,n+k} +\sqrt{n}\sqrt{n+k}N
  C_{n-1,n+k-1},
\end{equation}
which allows us to map each diagonal to a column vector $\vec{C}_k$, where $k$ denotes the 
distance from the principal diagonal.  
This explicit decoupling reduces the dimensionality of the problem from 2 to 1, making it easier to 
solve.  Specifically, we can now write the problem as
\begin{equation}
\frac{1}{\gamma}\frac{d}{dt}\vec{C}_k=\hat{A}[k]\vec{C}_k,
\end{equation}
which is much more tractable than Eq.~(\ref{numberbasis}), since it 
involves the derivative of a vector, rather than the derivative of a matrix.  

$\hat{A}[k]$ is an infinite matrix and is formally difficult to diagonalize. In Appendix A we 
show how to use numerical techniques to formulate an ansatz (that we go on to prove) for 
the diagonalization.  Solving the problem in diagonal space, and then transforming back 
to the original number space, the general solution is seen to be
\begin{widetext}
\begin{equation}
\label{generalC}
C_{n,n+k}=\sum_{j=0}^\infty
a_j\sqrt{\left(\begin{array}{c}k+n\\k\end{array}\right)}\left(\frac{N}{N+1}\right)^n\sum_{i=0}^n
(-1)^i\left(\begin{array}{c}n\\i\end{array}\right) N^{-i}
\left(\begin{array}{c}j\\i\end{array}\right)\frac{1}{\left(\begin{array}{c}k+i\\k\end{array}\right)}e^{-\gamma(j+\frac{k}{2})t},
\end{equation}
\end{widetext}
where $a_j$ represents infinitely many constants to be determined by the initial conditions.  
By using an infinite set of orthonormal initial conditions (again see Appendix A) we generate 
an orthonormal solution basis,
\begin{widetext}
\begin{eqnarray}
\nonumber
C_{n,n+k}[n_0,t]&=&\sqrt{\left(\begin{array}{c}k+n\\k\end{array}\right)\left(\begin{array}{c}k+n_0\\k\end{array}\right)}\left(\frac{N}{N+1}\right)^{n}\sum_{i=0}^n
\sum_{\nu=0}^{n_0}
(-1)^{i+\nu}\left(\begin{array}{c}n\\i\end{array}\right)
\left(\begin{array}{c}n_0\\\nu\end{array}\right)\nonumber \\
&\times& \frac{e^{(k+2)\gamma t/2}}{((N+1)e^{\gamma t}-N)^{i+\nu+k+1}}
\sum_{j=0}^{min(i,\nu)}\frac{\left(\begin{array}{c}i\\j\end{array}\right)\left(\begin{array}{c}\nu\\j\end{array}\right)}{\left(\begin{array}{c}j+k\\k\end{array}\right)}\left(\frac{N+1}{N}e^{\gamma
t}\right)^{j},
\label{partialsolution}
\end{eqnarray}
\end{widetext}
where $m\geq n$.  The techniques outlined above and shown in Appendix A can be applied to the case where $n\geq m$ by making the substitution 
$n=m+k$ into Eq.~(\ref{numberbasis}), yielding the almost identical orthonormal basis
\begin{widetext}
\begin{eqnarray}
\nonumber
C_{m+k,m}[m_0,t]&=&\sqrt{\left(\begin{array}{c}k+m\\k\end{array}\right)\left(\begin{array}{c}k+m_0\\k\end{array}\right)}\left(\frac{N}{N+1}\right)^{m}\sum_{i=0}^m
\sum_{\nu=0}^{m_0}
(-1)^{i+\nu}\left(\begin{array}{c}m\\i\end{array}\right)
\left(\begin{array}{c}m_0\\\nu\end{array}\right)\nonumber \\
&\times& \frac{e^{(k+2)\gamma t/2}}{((N+1)e^{\gamma t}-N)^{i+\nu+k+1}}
\sum_{j=0}^{min(i,\nu)}\frac{\left(\begin{array}{c}i\\j\end{array}\right)\left(\begin{array}{c}\nu\\j\end{array}\right)}{\left(\begin{array}{c}j+k\\k\end{array}\right)}\left(\frac{N+1}{N}e^{\gamma
t}\right)^{j}.
\label{partialsolutionm}
\end{eqnarray}
\end{widetext}
Equations~(\ref{partialsolution}) and (\ref{partialsolutionm}) together allow for a simple 
construction of a general solution to an initial density matrix $\rho[0]$ explicitly in terms 
of its elements represented in number space (whereas using only Eq.~(\ref{generalC}), 
one would have to derive the constants $a_j$ each time).  That construction is necessarily 
piecewise, since the cases $n>m$ and $n<m$ must be treated separately.  The final solution is
\begin{equation}
\label{rhosimp}
C_{n,m}[t]= \left\{
\begin{array}{c}
\sum_{z=0}^\infty C_{z,z+m-n}[0] C_{n,m}[z,t]\,\,\,n\leq m\\ \\
\sum_{z=0}^\infty C_{z+n-m,z}[0] C_{n,m}[z,t]\,\,\,n\geq m
\end{array} \right. .
\end{equation}
For a full derivation of this solution see Appendix A.  The infinite sums of Eq.~(\ref{rhosimp}) converge 
(assuming the initial density operator represents a quantum mechanically valid state) so that any desired 
accuracy can be achieved by the inclusion of sufficient terms.

Equation~(\ref{rhosimp}) simplifies greatly under certain special
conditions of interest. For example, if the initial state is a number state, we substitute $k=0$ into 
Eq.~(\ref{partialsolution}), and the solution takes the form
\begin{widetext}
\begin{equation}
\nonumber
P_n[t,n_0] =
\end{equation}
\begin{equation}
\left(\frac{N}{N+1}\right)^{n}\sum_{l=0}^n \sum_{i=0}^{n_0}
(-1)^{l+i}\left(\begin{array}{c}n\\l\end{array}\right)
\left(\begin{array}{c}n_0\\i\end{array}\right)\frac{e^{\gamma
t}}{(e^{\gamma t}(N+1)-N)^{i+l+1}}\sum_{\nu=0}^l
\left(\begin{array}{c}l\\\nu\end{array}\right)\left(\begin{array}{c}i\\\nu\end{array}\right)\left(\frac{N+1}{N}e^{\gamma
t}\right)^\nu.
\end{equation}
\end{widetext}
where the occupation probabilities $P_n=C_{n,n}$ are the diagonal entries of the density matrix. 
Janszky et.\ al.\ derived an equivalent result for these initial conditions in \cite{janszky90} 
by solving Eq.~(\ref{master}) in the Wigner representation.

A particularly important simplification we will consider is the case of number states decaying 
in a zero-temperature bath.  In this case the solution simplifies greatly:
\begin{equation}
\label{0tempsolnfock}
P_n[t]=\sum_{m=n}^{n_0} \left(\begin{array}{c}n_0 \\ m\end{array}\right) \left(\begin{array}{c}m \\ n \end{array}\right) (-1)^{n + m} e^{-\gamma m t},
\end{equation}
which is in agreement with Lu \cite{lu89}, who pioneered the idea of changing to the number basis, but did so only for the zero-temperature case.

Serafini et.\ al.\ derived a solution to a more general form of Eq.~(\ref{master}) which allows 
for the bath to be squeezed, or phase-sensitive\cite{serafini04}. Their solution, given an 
initial characteristic function $\bar{\chi}(x,p)$ is 
\begin{equation}
\label{serafinisolution}
\chi(x,p)=\bar{\chi}(xe^{-\gamma t/2},pe^{-\gamma t/2})e^{-(x\,p)\mathbf{\sigma}_\infty
\tiny
\left(\begin{array}{c}x\\p\end{array}\right)
\normalsize
(1-e^{-\gamma t})/2}
\end{equation}
where $\mathbf{\sigma}_\infty$ is the covariance matrix of the most general single mode Gaussian reservoir.  
Although formally able to solve the same class of equations as
our solution, it is numerically simpler to compute the dynamics of measures of non-classicality 
using our number-space representation, rather than working in characteristic space and calculating the
number-space entries of the density matrix from the characteristic function at each time step.

Given that our solution is already represented in number space, it is well-suited to comparison 
of state tomography experiments.  One such experiment was carried out by Brune et.\ al.\ \cite{brune08}, 
who used rubidium atoms in Rydberg states 
excited to the principle quantum number 50 and 51, and a pulsed microwave source to 
excite Fock states in a high-Q superconducting cavity.  They studied the behavior of the Fock
states as they decohered in a bath with mean-photon number $N=0.06$ using
QND (quantum non-demolition) measurements.  We found that our analytical
solution agreed extremely well with their results that present the
population of the various number states as a function of time. 

\section{Measures of non-classicality}
In order to quantify how non-classical a given state is, it is necessary
to first define what constitutes a classical state.  One 
definition of a classical state is one that exists as a possible solution 
in a classical dynamical system (even if the state satisfies the laws of quantum
mechanics as well). The formal answers to this question start from this
intuition, with the work of Klauder, Glauber and Sudarshan for example
in considering the coherent states in quantum 
optics\cite{klauder60,glauber63,sudarshan63}; it was 
later shown\cite{titulaer65} that coherent states could be thought of as 
classical states. The formal criterion provided for a state to be classical was 
that its Glauber-Sudarshan P-representation be positive definite and no more 
singular than a delta function.  

Another way of determining classical states comes from the intuition developed 
via the work on classical-quantum correspondence, particularly as it pertains to 
classically chaotic systems (see for example e.g. early definitive work by 
Berry, and later work represented by that of 
Wilkie et.\ al.\ \cite{berryjpa77,berrylond77,wilkie97a,wilkie97b}). 
Berry pioneered the understanding that quantum-classical correspondence
was best analyzed in the the Wigner-Weyl representation, which he was
able to prove is the only phase-space formulation of quantum mechanics 
that approaches the classical limit of Liouville phase-space mechanics.  
In this case, the criteria for a state to be considered classical are that 
its Wigner representation be continuous, smooth, positive definite, and 
normalized. Coherent states are gaussians in the Wigner representation that 
minimize the uncertainty principle, and satisfy all the properties of a 
classical probability distribution. They are indistinguishable from a 
classical probability distribution that starts with a minimum uncertainty.

A physically-argued justification that coherent states are the quantum 
analogues of points in phase-space also comes from Zurek et.\ al.\ \cite{zurek93}, 
who showed that coherent states are the most stable in a thermal bath. 
Specifically they studied decohering pure states weakly coupled to a thermal 
bath, and showed that the coherent states produced the least entropy as they 
decayed. It has also been established elsewhere that the coherent states are 
the only pure states that are classical \cite{hillery85}.  In phase-space this argument
takes the form of a theorem that the only positive-definite pure state Wigner functions 
correspond to Gaussian states.

The most common mixed states which fit all the above descriptions of classical states
are thermal states -- whether centered at the origin or displaced -- and have come to 
be a standard addition to the set of classical states when studying relative measures 
(for example see \cite{dodonov03, marian02, marian03}).  Thermal states centered at 
the origin are the steady state solutions to Eq.~(\ref{master}), are defined as
\begin{equation}
\rho_{th}=\frac{1}{N+1}\sum_{n=0}^\infty \left(\frac{N}{N+1}\right)^n,
\end{equation}
and represent a thermal distribution of number states, with $N$ defined, as in 
Eq.~(\ref{master}), to be the mean number of thermal photons in the bath.  The displaced
thermal states are defined using the displacement operator 
$\hat{D}(\alpha)=\exp{\alpha \hat{}a^\dagger-\alpha^* \hat{a}}$:
\begin{equation}
\hat{D}(\alpha)\rho_{th}\hat{D}^{-1}(\alpha).
\end{equation}
However, as we will discuss in Section IV, including the displaced thermal states in the set of classical states
is not sufficient for studying the dynamics of the decohering harmonic oscillator, and we will 
present a novel addition to the basis of classical states.

We note several caveats here: 
(a) First, although the quantum optics and the phase-space perspectives agree 
on the classicality of thermal states, there is an important way in which 
they disagree.  Specifically, the phase-space perspective considers {\em all} 
gaussians to be 
classical even though this includes the squeezed coherent states, since the 
criteria that define a valid classical Liouville distribution are invariant 
under squeezing. However, squeezed states are considered non-classical in 
quantum optics since such states exhibit sub-Poissonian photon 
statistics.  
Our perspective is based in phase-space and we do not consider squeezed
states as a result. 

(b) Of course these classical states -- the coherent states and the displaced thermal states are 
indeed quantum mechanical states in that they satisfy the appropriate quantum equations and 
evolve correctly in the quantum systems.  However, their properties are indistinguishable from those 
of classical Liouville probability distributions with a requirement of minimum uncertainty. 

(c) It is also important here to stress that we are only considering unipartite systems.  If 
for example two coherent states are entangled (creating a cat state), or a coherent 
state is used to entangle two other particles, that coherent state can no longer be considered as 
a single entity, and the total multipartite system would certainly not be considered classical.

Having established a set of classical states, it is then possible to classify how non-classical another state is 
by considering how different the given state is from the set of classical states. We will consider three of 
the standard ways of quantifying this distance between an unknown state and the set of classical states.  

These are relative measures, comparing unknown states to those known to be classical.  
These are formally well-defined. However, as we see below, the
non-classicality of a state is defined in terms of the infimum (or
in the case of classicality measures, the supremum) of the distance from {\em all}
classical states. Given that it is impossible to search through
this infinite set of all classical states, there is a certain amount of
ambiguity about these relative measures.  Progress can be made using an
incomplete basis, however, as has been previously achieved, and in
particular in some cases upper and lower bounds on the non-classical distance can be 
established \cite{hillery87}.

Other ways of quantifying non-classicality can be understood as absolute
measures in that they depend only on inherent properties of the state under 
consideration, and so avoid the problem of defining a set of classical states.  
Perhaps the simplest and most well-known examples of this type of measure is the purity, 
$\textrm{Tr}\rho^2$, or its close relative, the 2-entropy, $\textrm{Log}[\textrm{Tr}\rho^2]$.  
For example, as mentioned in passing above, Zurek et.\ al.\ \cite{zurek93}  studied dynamics 
by calculating the entropy (specifically the 2-entropy) generated 
by the decohering pure states, and offered further
evidence that the coherent states are the most classical pure states.  
Marian et.\ al.\ have also used the same measure to show that in most cases as 
Fock states decay in the Markovian bath, they reach a point of maximal mixing, after which 
time the purity or 2-entropy increases again.  This result indicates
that while the purity can be a reasonable rule of thumb in the static case,
it does not hold up in the dynamical case as a sensible measure of non-classicality.  
This is because for an open system it does not conform with intuition that non-classicality 
should decay to a minimum, and then increase again under the effects of decoherence. 
Further, coherent states, which have been defined as the most classical states of all by 
many different authors, have unit (maximum) purity.  Dodonov et.\ al.\ make similar arguments
against using purity as a measure of non-classicality \cite{dodonov03}.

Many absolute measures involve representing the uncharacterized state in phase space and 
measuring some property of that representation.  Paavola et.\ al.\ , who
studied dynamical systems using nonclassical measures, present five such measures and 
use them to study the decoherence of cat states \cite{paavola11}.  They consider how long it 
takes for the Wigner function to become fully positive, the time it takes for the 
Glauber-Sudarshan function to become fully positive, the time it takes for the interference 
fringe in the Wigner representation to disappear, the Vogel criterion, and the Klyshko criterion.  
They show that of these five measures, all but the interference fringe technique show that 
the cat state becomes classical in a finite amount of time, and that those times are all 
on the same order of magnitude.  Since they have already shown these four approaches to give 
similar results, we will only consider one of the measures in this paper.  The only other paper 
studying dynamics of non-classicality that we are aware of also studied the negative volume of
the Wigner function of cat states \cite{dodonov11}.  In particular, that paper presented analytical 
formulas for when the Wigner functions of cat states become fully positive.

With this as background, the four measures we present results on below are
\begin{enumerate}
\item The first is a measure discussed in detail by Hillery, being
defined as the infimum of the trace norm of the difference between the quantum state and 
the set of all classical states $\rho_{cl}$ \cite{hillery87}: 
\begin{equation}
\label{hillerymeasure}
\eta_{H}[\rho]=\inf_{\rho_{cl}}||\rho-\rho_c||
\end{equation}
where $||A||=\textrm{Tr}[\sqrt{AA^{\dagger}}]$ is the trace norm of a matrix.  
Hillery used this measure to prove that there will always be a finite distance 
between non-classical states and classical states.  He also used it to study 
the harmonic oscillator eigenstates, and proved that $\eta_H$ is bounded from above 
and below by
\begin{equation}
1-\gamma_n\leq\eta_H\leq2\sqrt{1-\gamma_n}
\end{equation}
and that $\gamma_n$ must decrease at least as quickly as $n^{-1/2}$ for large $n$.  
This result simply rests on the existence of a 
classical basis, and does not require its construction.  In Section IV. we use a 
partial classical basis to show that non-classicality in fact increases monotonically 
and asymptotically to the maximum value predicted of 2.  Though it is not 
immediately evident from Eq.~(\ref{hillerymeasure}), the maximum value 
of $\eta_H$ is in fact 2, meaning that increasing number states approach the 
maximal value of non-classicality.

\item The second measure, the Bures distance, uses the Uhlmann fidelity, given by 
$B[\rho_1,\rho_2]=\textrm{Tr}[\sqrt{\sqrt{\rho_1}\rho_2\sqrt{\rho_1}}]$, which was 
shown to be the density operator generalization of $|\langle \psi_1|\psi_2\rangle|$ 
by Uhlmann \cite{dodonov03}. The actual Bures distance is calculated using \cite{marian02}
\begin{equation}
\label{buresmeasure}
\eta_B=\inf_{\rho_{cl}}\sqrt{2(1-|B[\rho,\rho_{cl}]|)}.
\end{equation}
This measure was used to determine analytically the non-classicality of squeezed, displaced 
thermal states \cite{marian02}, and later the non-classicality of two
such states when entangled \cite{marian03}.  In Section IV. we show that this measure 
agrees well with the Hillery measure in the case of the number states.  The only real 
difference, that $\eta_B$ increases asymptotically to $\sqrt{2}$ instead of 2, is 
due simply to its construction.  Eq.~(\ref{buresmeasure}) clearly shows that the maximum 
value $\eta_B$ can attain is $\sqrt{2}$.

\item The third measure (the Dodonov measure), is a measure of classicality, rather than 
non-classicality, as it computes how close, rather than far, a state is to the classical 
basis.  It is presented in \cite{dodonov03}, and is a modified form of the overlap of 
density operators: 
\begin{equation}
\eta_D[\rho]=\sup_{\rho_{cl}}\textrm{Tr}[\rho'\rho_{cl}']
\end{equation}
where $\rho'=\frac{\rho}{\sqrt{\textrm{Tr}[\rho^2]}}$ is the density operator renormalized 
by its purity.  
The purpose of this renormalization is to make it easier to compare mixed states.  
Notice that for mixed states Tr$[\rho_1'\rho_2']=1$ if $\rho_1=\rho_2$, whereas 
Tr$[\rho_1\rho_2]<1$ in the same case.  Thus, the renormalized overlap 
allows one to easily see how similar two density operators are, regardless of the 
degree of mixedness.  Dodonov et.\ al.\ used this measure to show that the classicality 
of the number states decreases monotonically and asymptotically to 0.  Again this 
means that increasing number states approach the maximal value of non-classicality.  
Specifically they showed that $\eta_D=e^{-n}n^n/n!$, which, using the Sterling 
approximation, reduces to $\eta_D\approx(2\pi n)^{-1}$ for large n.

\item The measure we will consider from the family of absolute measures is the 
negative volume of the Wigner function, or negativity,
\begin{equation}
\eta_W=\int_{-\infty}^\infty \int_{-\infty}^\infty (|W[x,p]|-W[x,p]) \textrm{d}x\textrm{d}p,
\end{equation}
which is particularly appealing because it is easily calculated using
numerical integration.  It is also consistent with the intuition resulting from the 
study of quantum-classical correspondence in phase-space following Berry.
Kenfack et.\ al.\ have already used $\eta_W$ to study the harmonic
oscillator eigenstates (for the closed, static problem) Their results agreed very well 
with those of Dodonov et.\ al.\ in that Kenfack et.\ al.\ found that $\eta_W$ increases 
like $\sqrt{n}$ \cite{kenfack04}. 
\end{enumerate}

We now present results using these measures, starting with the previously well-studied (see e.g. \cite{hillery87,dodonov03}) base-line case
where there is no coupling with the environment (which can alternatively be
understood as the initial condition for all the cases).

\section{Results}

\subsection{No Environmental Coupling}

As discussed in section III, the Hillery and Dodonov et.\ al.\ measures have been previously used to 
show that the non-classicality of Fock states increases with increasing $n$ in the absence 
of any environmental coupling.  Dodonov et.\ al.\ derived an analytical expression for this 
increase using the classical basis composed of the displaced thermal states.  
In studying this problem we found and present here an improvement upon
both these basis, which minimizes $\eta_D$ further (recall that this
measure seeks the infimum over all classical states), and further 
maximizes $\eta_H$ and $\eta_B$ (recall that these measures seek the 
supremum over all classical states).  

We discovered this new basis -- particularly of relevance to decohering
states -- through our own numerical experiments as well as from studying previous results
on the behavior of initial number states\cite{knight96}. 
What has been observed is as follows: the Wigner functions for the number 
states start as Laguerre polynomials in phase-space, with multiple
non-classical fringes. As the states evolve in the presence of the 
environment, they lose their initial fringes until they become slightly 
broadened versions of the classical microcanonical states. That is, they 
become positive-definite distributions which are sharply peaked in 
energy space on the appropriate classical energy, and are otherwise evenly 
distributed along the classical orbit. The classical microcanonical state 
at a given energy $E$ is, of course, $\delta (E -H(p,q))$; the states we are talking about
are not quite as singular as these delta functions in energy in having a 
slight spread in energy. 

These 'thermally-broadened' microcanonical states are a natural choice
as classical states which might be closest to these quantum eigenstates.
In studying quantum-classical correspondence averaging over neighboring
states has been shown to be necessary, yielding states similar to these
thermally broadened micro-canonical states \cite{berryjpa77,berrylond77,wilkie97a}.  
Irrespective of the intuition, as we show below, empirically these
states work very well indeed.

We represent these states by
\begin{equation}
\rho^+ =\sum_{n=0}^\infty a_n\rho_n^+,
\end{equation} 
where $\sum_{n=0}^\infty a_n=1$ and $a_n\in [0,1]$, and $\rho_n^+$ represents a 
number state that has evolved according to Eq.~(\ref{master}) to a time 
$t_*=\frac{1}{\gamma}\log{\frac{2N+2}{2N+1}}$ \cite{dodonov11}, which guarantees that it
is positive definite in the Wigner representation. 
This is an extremely large set and is impossible to explore completely.  
For our purposes, we found that the most useful subset of $\rho^+$ is 
\begin{equation}
\rho_\nu^+=(x+1-\nu)\rho_x^++(\nu-x)\rho_{x+1}^+
\end{equation}
where $x=\textrm{Int}[\nu]$ is the truncated integer of $\nu$, and $\nu$ 
varies continuously.  For example by this 
definition $\rho_{2.9}^+=.1\rho_2^++.9\rho_3^+$.

\begin{figure}[h]
\caption{For the static (or closed/non-interacting) case we show
different measures of non-classicality for eigenstates of the harmonic
oscillator. Specifically we show the \textbf{(a)} Hillery Distance, $\eta_H$,
\textbf{(b)} Bures Distance, $\eta_B$, and \textbf{(c)} Dodonov Overlap,
$\eta_D$, each computed for three different classical bases, and \textbf{(d)} negativity, $\eta_W$.}
\label{fig:staticcomparisons}
\subfigure[]{
\includegraphics[width=3in]{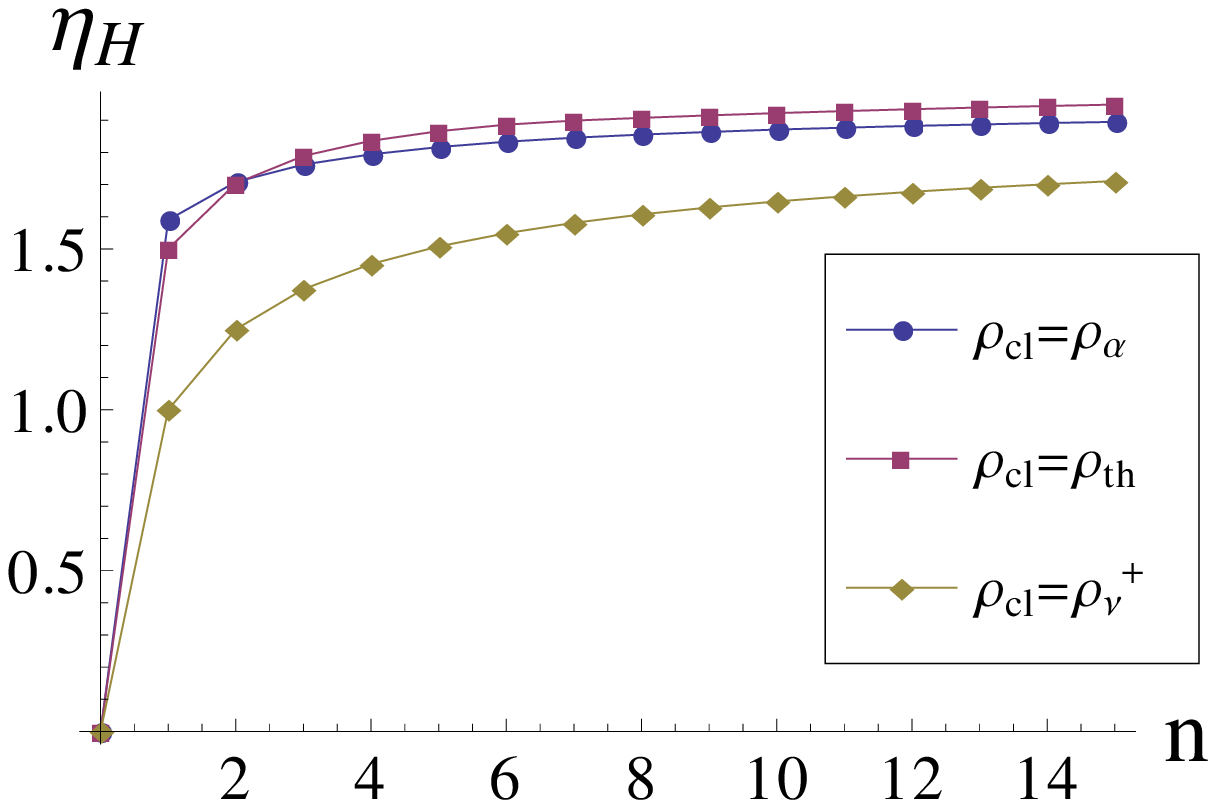}
\label{fig:Hstaticcomparison}}
\subfigure[]{
\includegraphics[width=3in]{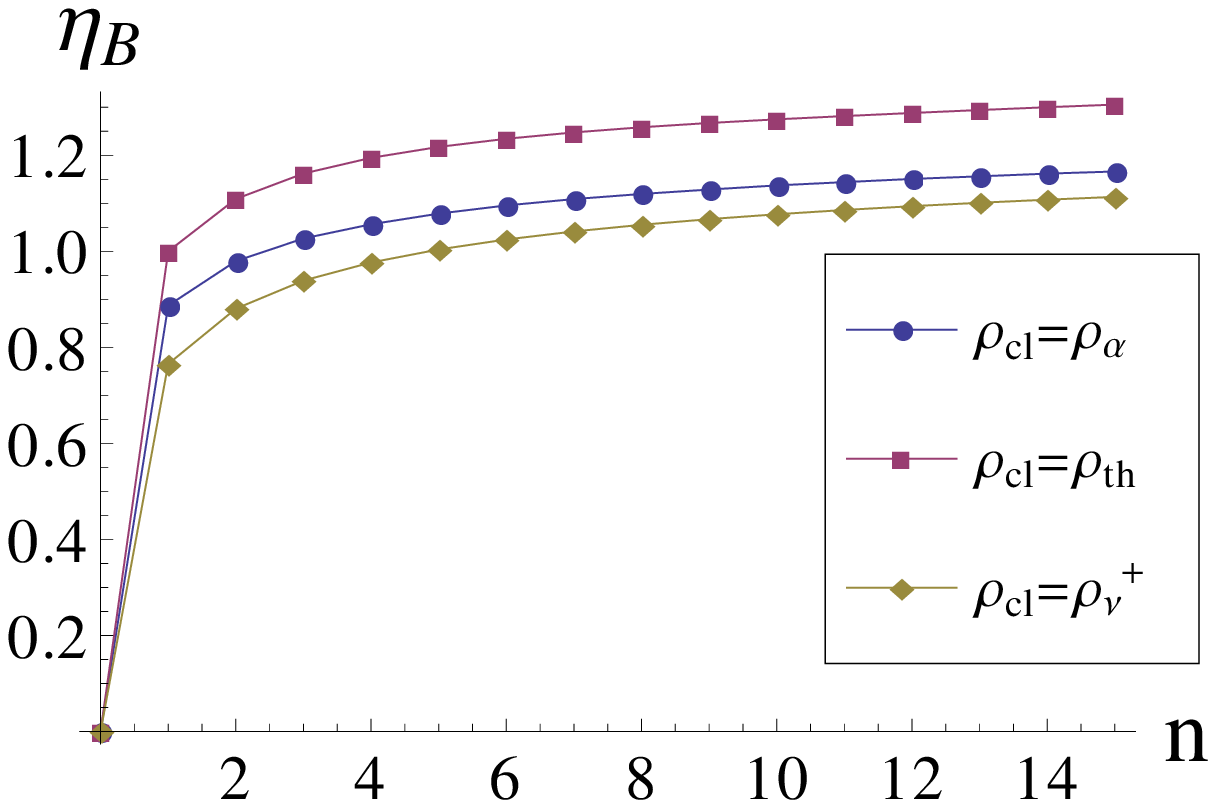}
\label{fig:Bstaticcomparison}}

\subfigure[]{
\includegraphics[width=3in]{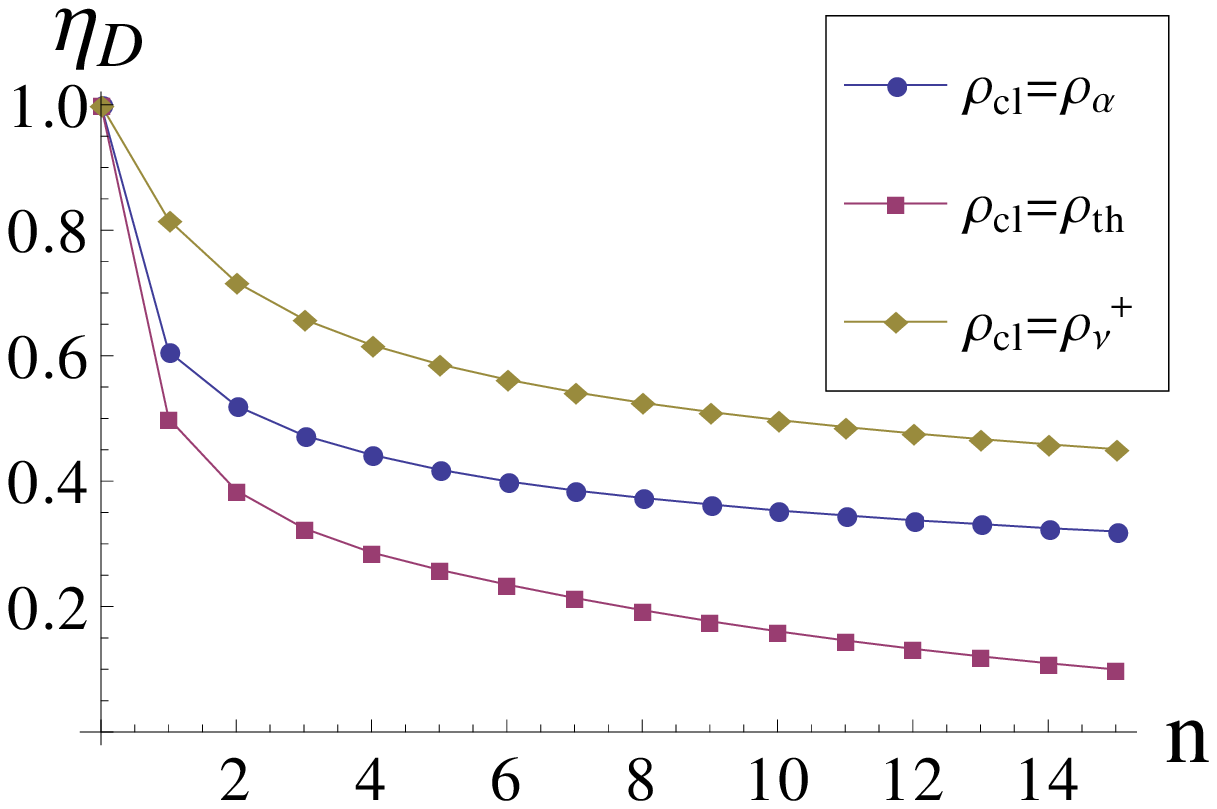}
\label{fig:Dstaticcomparison}}
\subfigure[]{
\includegraphics[width=3in]{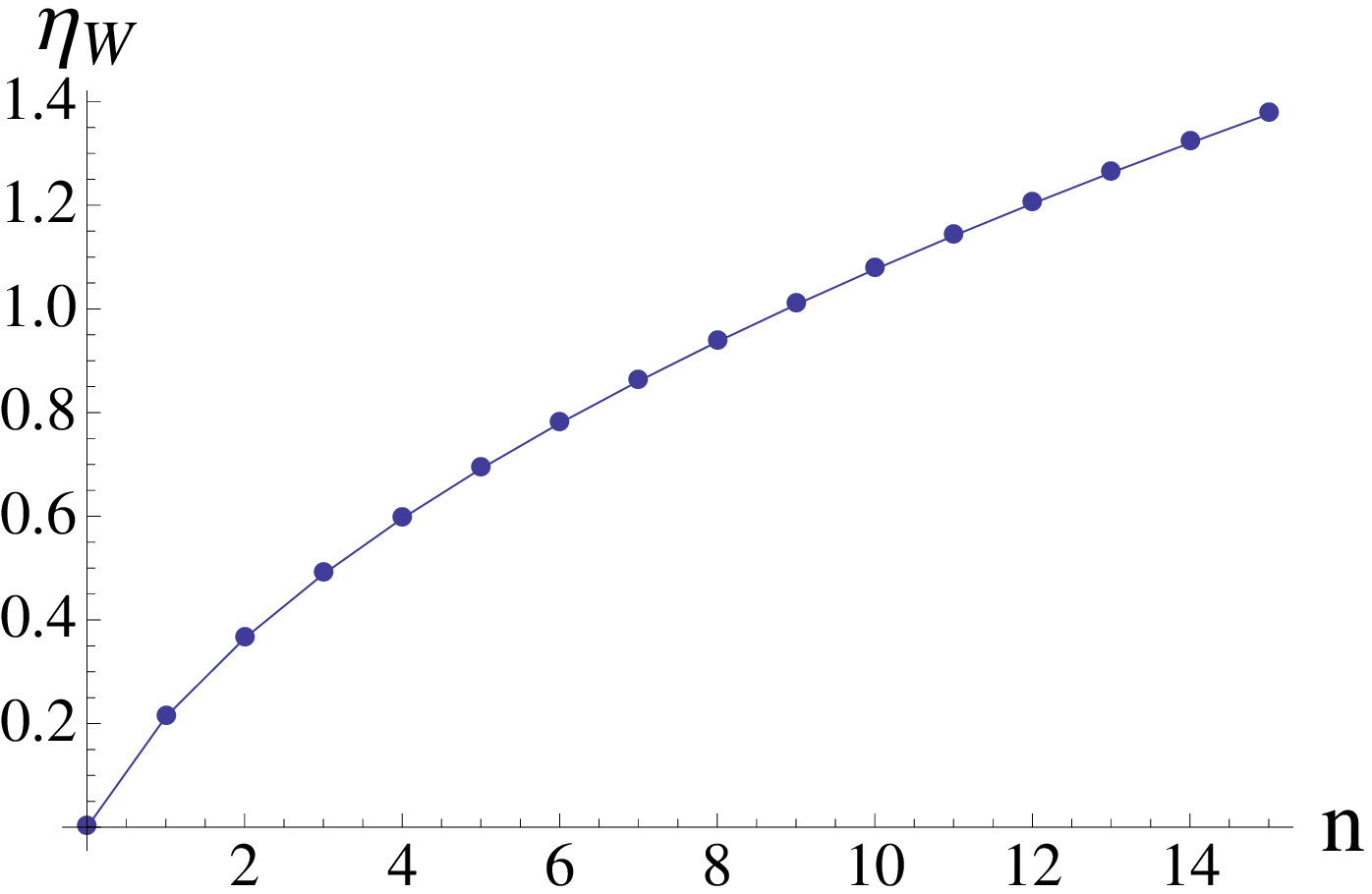}
\label{fig:Wstaticcomparison}}
\label{fig:static}
\end{figure}

In Fig.~\ref{fig:static} we show the results of computing these measures of
non-classicality for eigen-states of the harmonic oscillator using
multiple measures, and using three different classical bases: the set of 
all coherent states, the set of all thermal states, and the set $\rho_\nu^+$. 
We note that although clearly $\rho_\nu^+$ outperforms the other bases, it does not 
provide any new insight into the behavior of the system.

\subsection{Zero Temperature}

The results of the previous section seem to violate the correspondence
principle on the face of it, since they show higher $n$ states as more 
non-classical in contradiction to the rule of thumb that higher number 
states should be less non-classical.
This apparent paradox has in fact been well known for decades: that
higher $n$ states display more non-classicality -- via rapidly
oscillating fringes, for example -- in violation of the correspondence principle 
intuition. It has been previously shown that the correct way to approach
this issue is by averaging the Wigner function over a small energy
spread or phase-space spread (usually evoked as a resolution limit).
When such averaging is done for the Fock states it appropriately smooths the 
fringes, or oscillations, of the number states
\cite{berryjpa77,berrylond77}.  The same smoothing effect occurs explicitly when 
studying dynamics.  As soon as environmental coupling is considered, the number 
states evolve over time into states resembling the micro-canonical states 
(in this case the `averaging' as being done by the environment).  
Higher $n$ states are much more sensitive to decoherence than lower $n$
states, which means smoothing occurs much more rapidly for higher $n$
states, and yields the appropriate correspondence principle intuition.

Another way of understanding this idea, that the correspondence
principle requires averaging (either mathematical or environmental),
starts from the notion that the behavior of closed systems is singular
and non-physical, in that states in such systems evolve completely independent 
of their environment; they do not even interact with the vacuum.  This
behavior does not survive when even the smallest amount of interaction
with the environment is introduced.

\begin{figure*}
\caption{
These plots show the evolution of non-classicality for the
zero-temperature case. Each curve represents a different snapshot in time of
non-classicality plotted as a function of initial eigenstate $n$. 
We show this for the \textbf{(a)} Hillery distance, \textbf{(b)} 
Bures distance, \textbf{(c)} Dodonov overlap, and \textbf{(d)} negativity.}
  \begin{minipage}[b]{5.5in}
    \begin{flushleft}
\subfigure[]{
\includegraphics[width=2.6in]{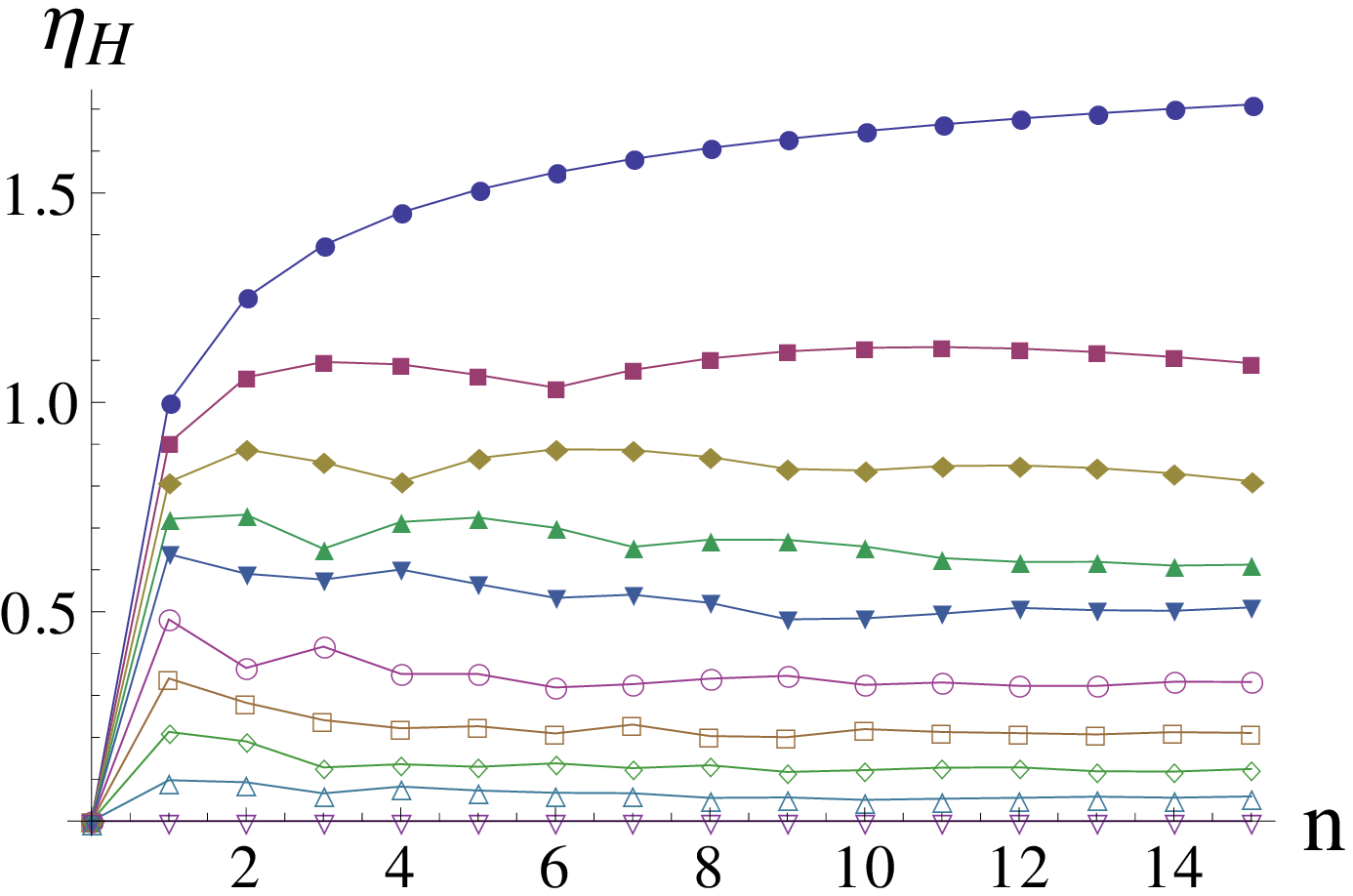}
\label{fig:qpeakH}
}
\subfigure[]{
\includegraphics[width=2.6in]{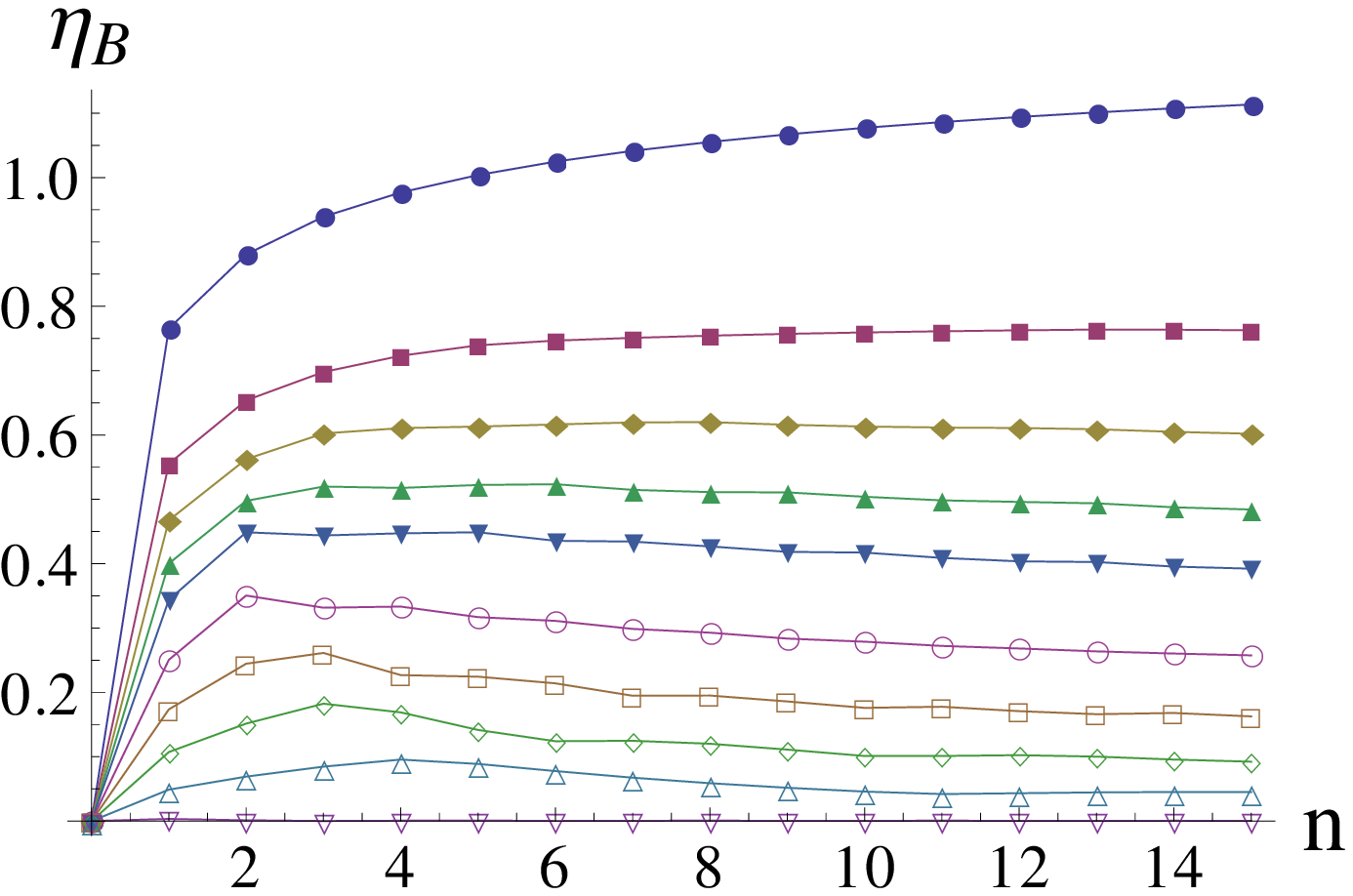}
\label{fig:qpeakB}
}

\subfigure[]{
\includegraphics[width=2.6in]{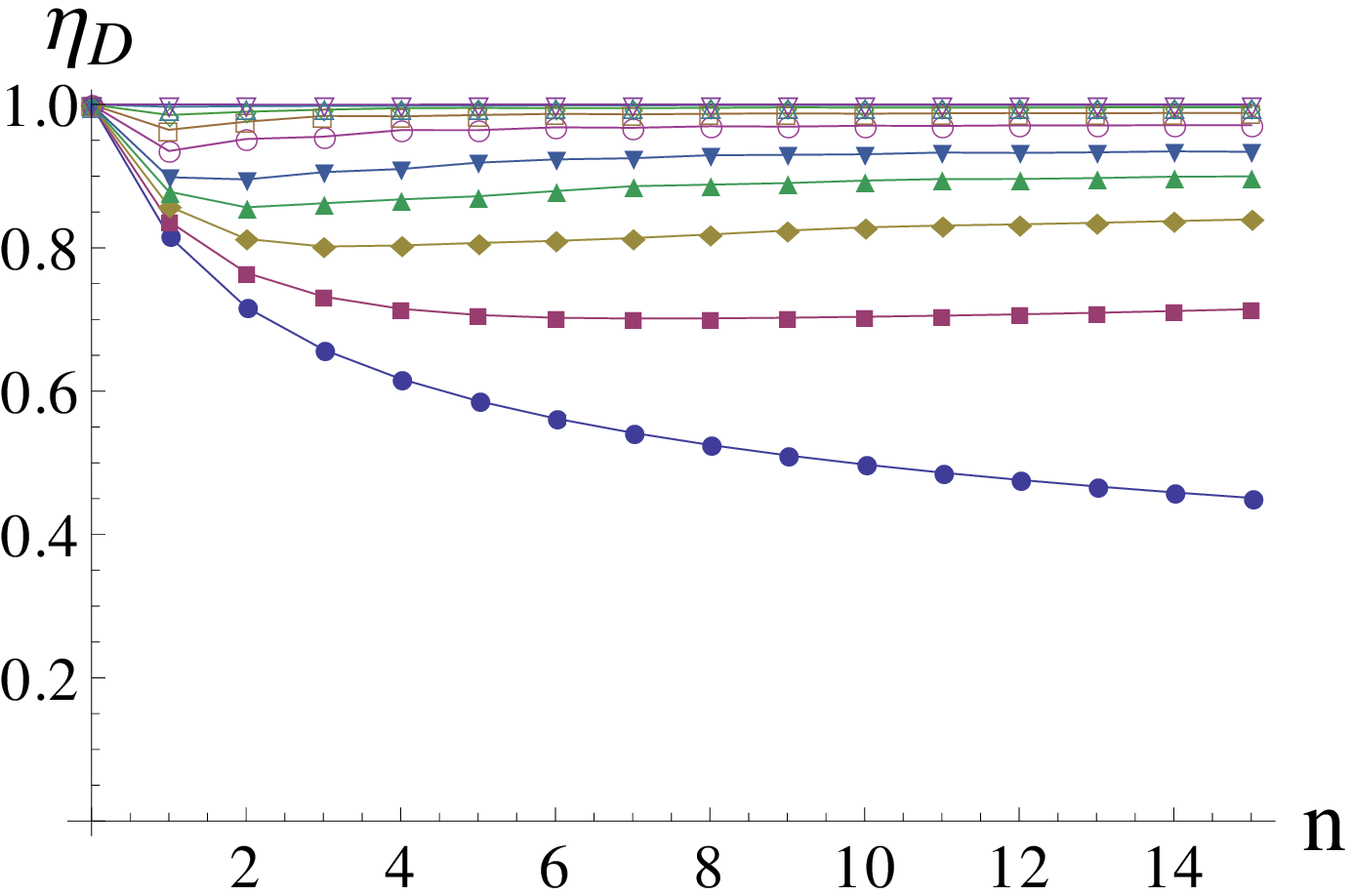}
}
\subfigure[]{
\includegraphics[width=2.6in]{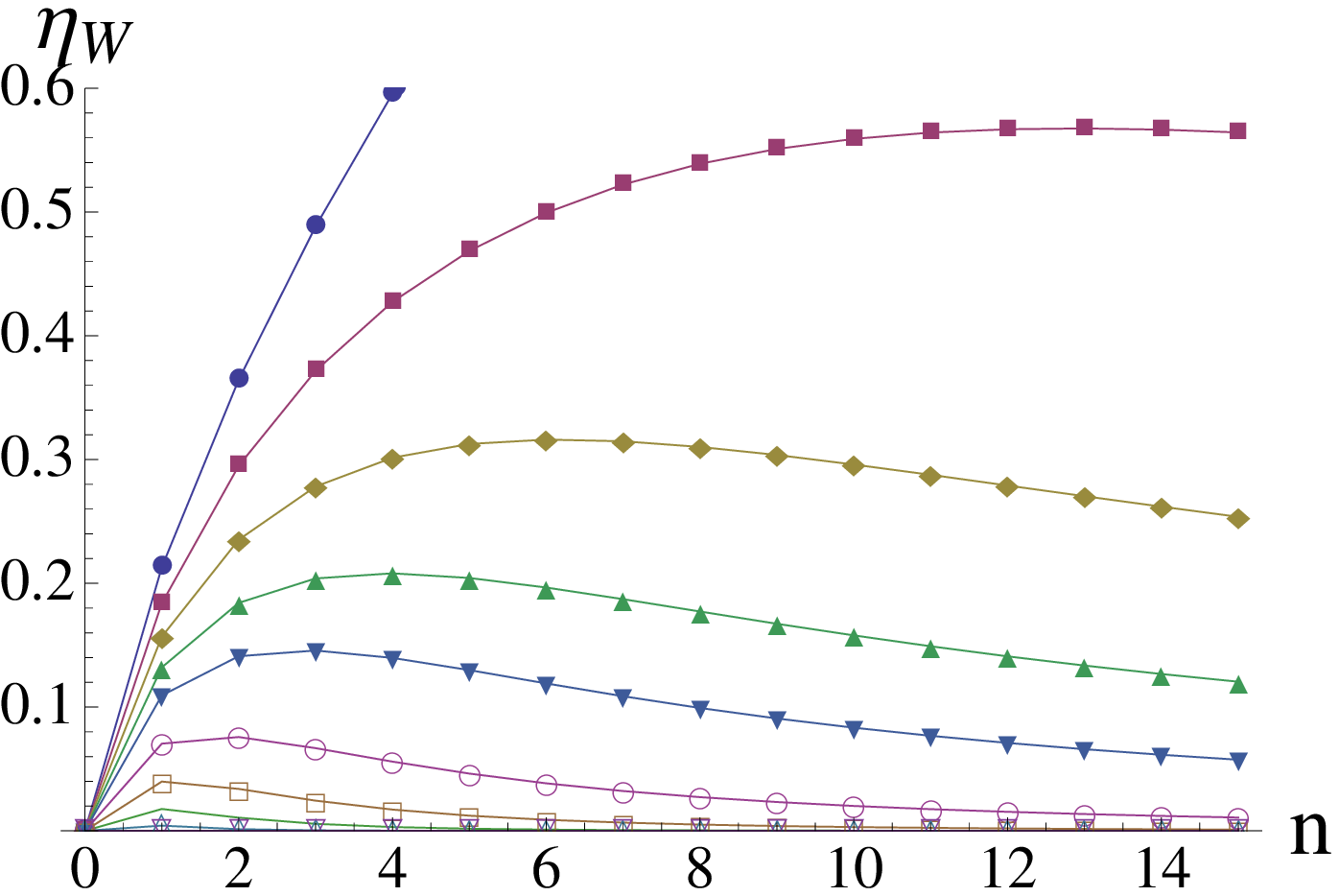}
\label{fig:etawvsn}
}      
    \end{flushleft}
  \end{minipage}
  \vspace{0.5cm}
  \begin{minipage}[b]{.9in}
    \begin{flushright}
      \subfigure{
        \includegraphics[width=.9in]{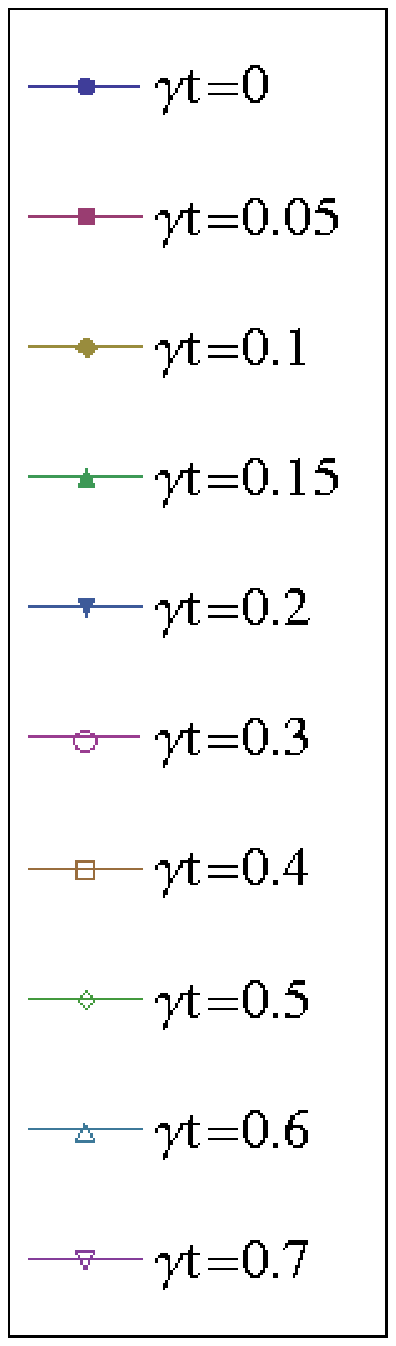}
        }
    \end{flushright}
  \end{minipage}
\label{fig:qpeak}
\end{figure*}

Specifically, for example, by introducing a zero-temperature bath into the 
problem, the behavior of the system changes markedly from that displayed in 
Fig.~\ref{fig:static}.  Using the expanded classical basis outlined above, 
each of the three relative measures, along with the absolute negativity 
measure, yield results in agreement with the correspondence principle (see 
Fig.~\ref{fig:qpeak}). That is, all of the non-classicality curves shown for 
finite $\gamma t$ peak, and then asymptotically decrease to $0$ with 
increasing $n$ (although this is harder to see for the Hillery and Bures
measures.  The new basis turns out to be critical in detecting 
this.  Without it (that is, using only the displaced thermal staes), 
according to the relative measures the non-classicality of the decohering 
eigenstates shows a monotonic \emph{increase} for all $\gamma t$ curves. 

\begin{figure}
\caption{The decrease in non-classicality as a function of time for the $1st$, $4th$ and $7th$ Fock states as measured 
by the \textbf{(a)} Hillery distance, \textbf{(b)} Bures distance, \textbf{(c)} Dodonov overlap, and \textbf{(d)} negativity.}
\subfigure{
\includegraphics[width=3in]{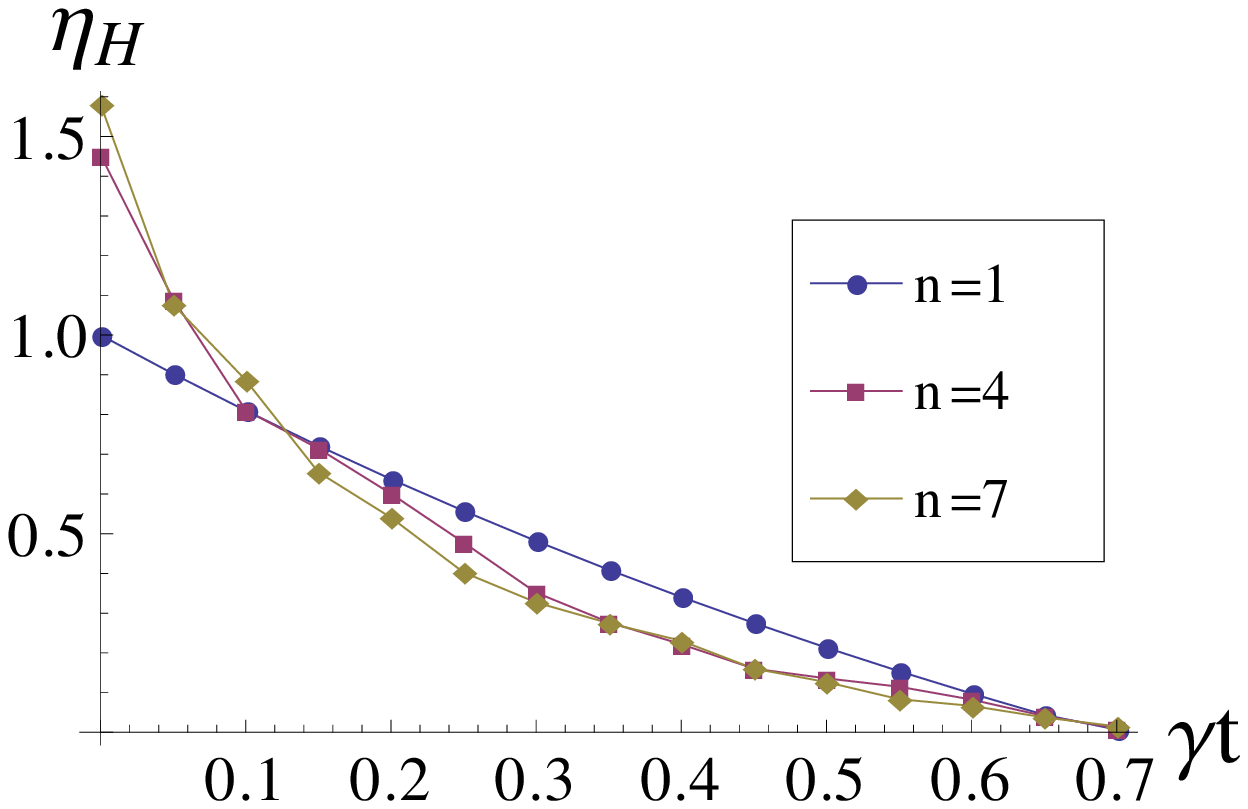}
\label{fig:H147}
}
\subfigure{
\includegraphics[width=3in]{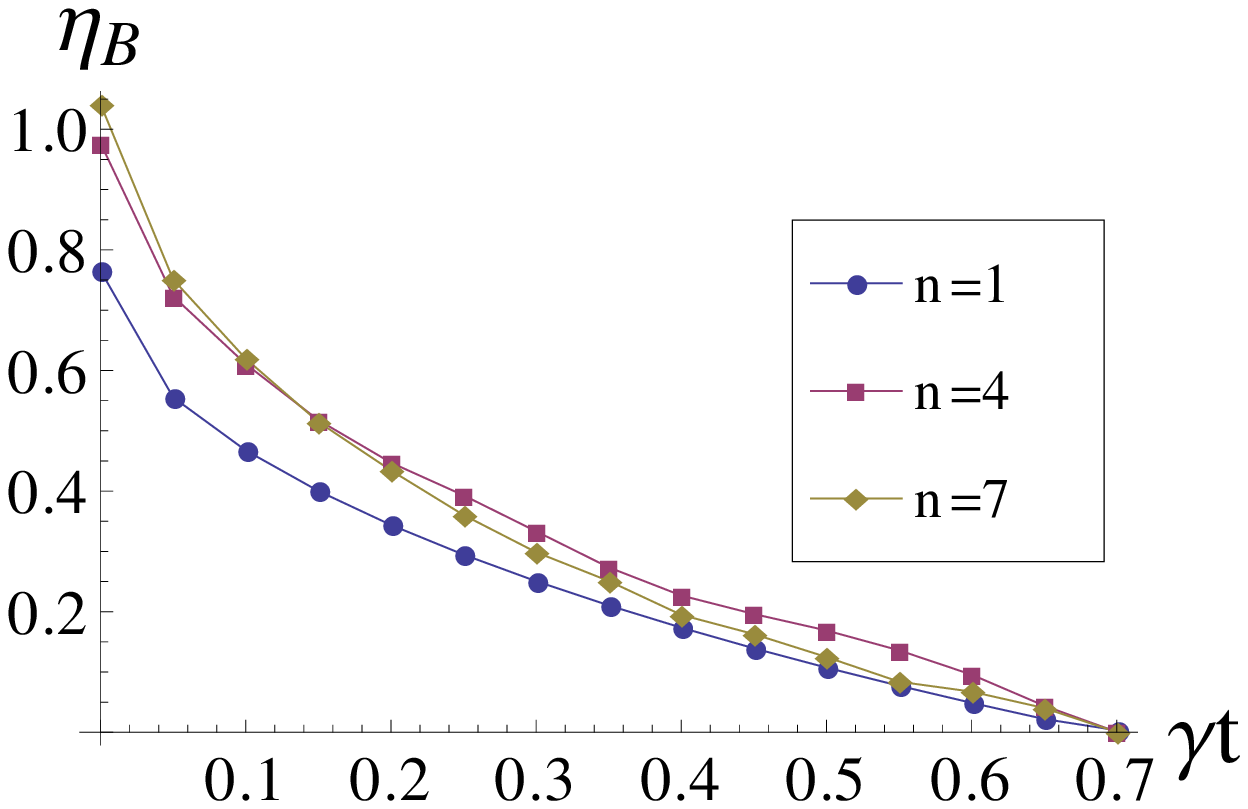}
\label{fig:B147}
}

\subfigure{
\includegraphics[width=3in]{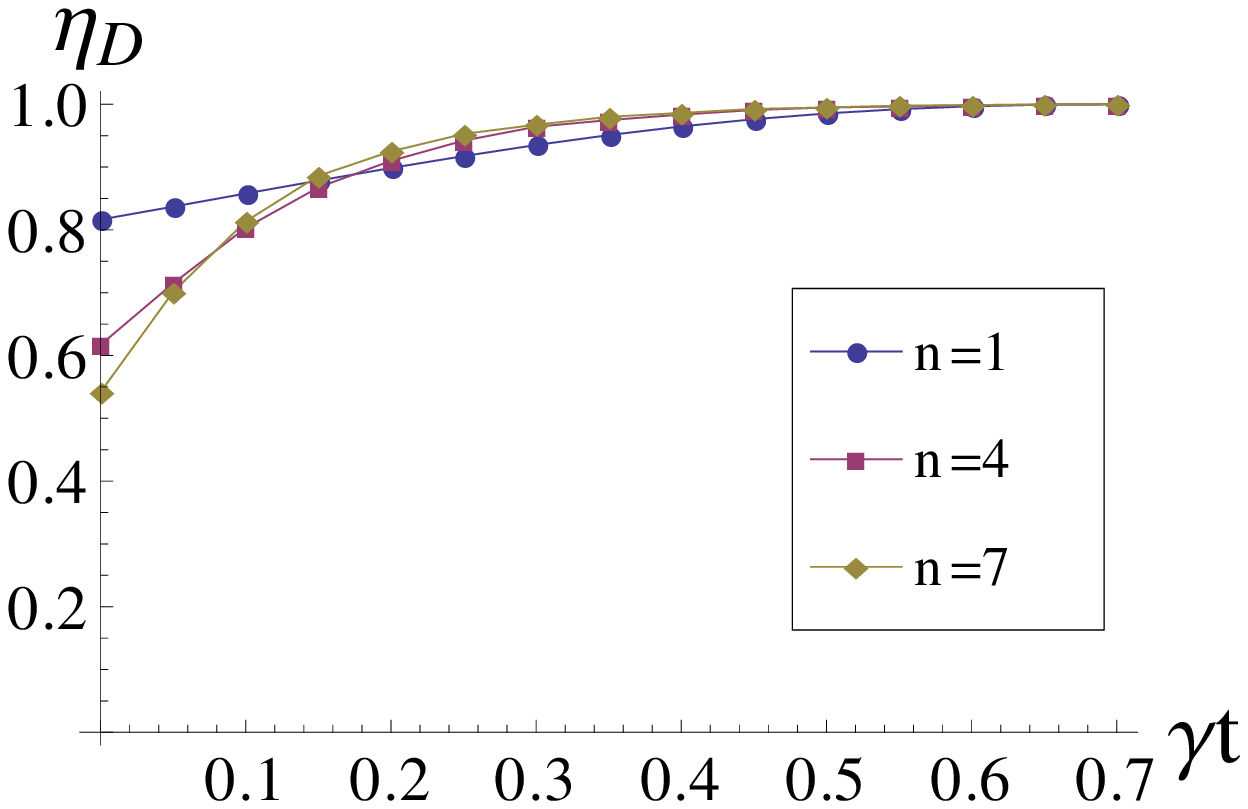}
}
\subfigure{
\includegraphics[width=3in]{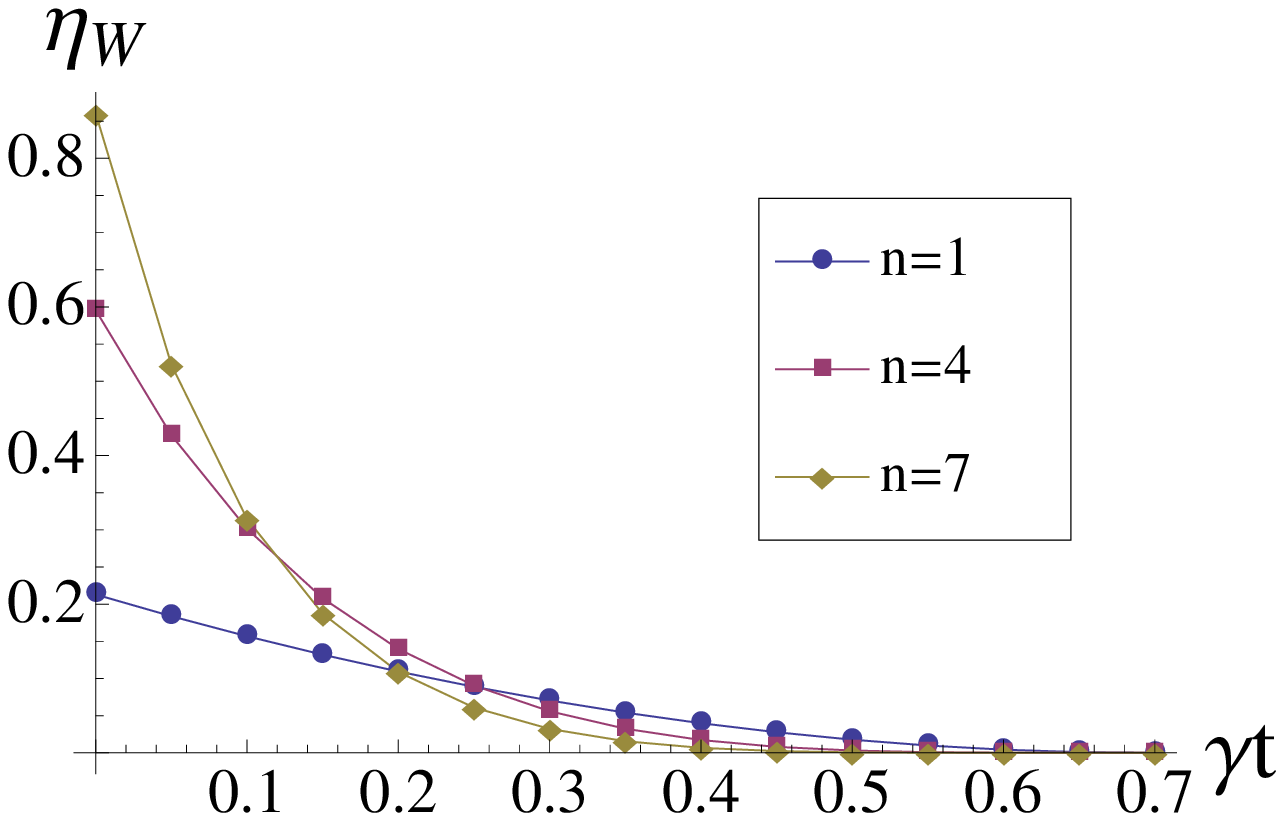}
}
\label{fig:qvst}
\end{figure}

Though each of the measures show the same general result, each one 
differs somewhat on the details of the behavior.  According to (c) the Dodonov 
overlap and (d) negativity, this non-monotonicity is relatively 
straightforward.  Each state decays smoothly, and the curves for each
state each other state only once, so that the non-classicality peak 
decreases in $n$ monotonically.  However, the (a) Hillery and (b)Bures distances 
tell a more complicated story.  As can be seen in Fig.\ \ref{fig:H147}, 
individual eigenstates do not decay smoothly as a function of time according 
to the Hillery distance; specifically the $4th$ and $7th$ excited states exhibit 
multiple corners, giving rise to multiple crossovers.  This naturally leads 
to the more complex behavior seen in Fig.\ \ref{fig:qpeakH}.  The time curves 
in this case display not only a maximum value, but also local minima.  

The Bures distance shows individual eigenstates decaying more smoothly than 
the Hillery distance, though Fig.\ \ref{fig:qpeakB} does show occasional 
local minima.  The more unusual signature present in this measure is that 
many of the eigenstates never cross the decay curve of the $1st$ eigenstate, as 
seen in Fig.\ \ref{fig:B147}.  For this reason the non-classicality peak never 
reaches $n=1$.  The lowest value it reaches $n=2$, after which time it begins 
moving up again.  

These unexpected features may be explainable by the fact that we used a classical 
basis that is far from complete.  Ideally the Hilbert-space measures
rely on searching over the entire set of classical states, which is practically 
impossible.  Unfortunately this inability to represent the complete classical 
basis can lead to incorrect results.  Indeed, as we have mentioned, when using 
the coherent or thermal states as reference states the non-monotonic signature 
disappears altogether.  It may be reasonable to assume that the non-classicality 
peak would come to rest at $n=1$ according to the Bures distance as well if a 
more complete basis were used.  Similarly, it is arguable that the Hillery distance 
would decay smoothly for all harmonic oscillator eigenstates if a more complete 
basis were found.

Independent of these speculative arguments about these unusual features, 
all four measures show qualitatively similar behavior. That is, in all 
cases, an initially ($\gamma t=0$) monotonic dependence on quantum number 
transforms into non-monotonic behavior in the presence of decoherence. 

We comment here about the relative computational difficulty of the
different measures. We have just demonstrated that in the case of zero 
temperature negativity (the negative volume of the Wigner function) shows good 
agreement with the other three measures discussed.  However, comparing
these measures in more general cases is even more formidable challenging.
Specifically we would have to search even more broadly to discover the
best classical bases for each problem. Further, the zero temperature
problem is relatively convenient computationally speaking, because it deals 
with finite density matrices in the number representation.  The non-zero 
temperature leads to infinite density matrices in number space, and makes 
calculating non-classicality more and more expensive as temperature increases.  
For negativity, only one numerical integral must be calculated at every
time step, whereas for the other measures hundreds of values must be calculated 
and compared at each time step, so the difficulty of the problem scales much more
quickly.  Given the computational difficulty of the relative measures, and the
uncertainty of finding adequate bases for further study, we choose to
use solely the negativity as a measure of non-classicality for the rest of 
our studies.

For those who are still suspicious of using negativity as a measure of
non-classicality, these results may well seem unsatisfying.  However, we 
would like to remind the reader that the only other two studies of dynamics
we found \cite{paavola11,dodonov11} dealt solely with phase-space measures.
We have attempted to validate this approach by comparing the measures 
where we found it possible, just as Kenfack et.\ al.\ did before \cite{kenfack04}
by considering the static case.  In addition, we point out a further pragmatic
argument for considering negativity to be useful in quantifying 
non-classicality, which comes from somewhat recent discoveries in the field of 
quantum computation that deal with a discrete analogue 
of the Wigner function.  Cormick et.\ al.\ have 
shown that the only non-negative pure states in this representation are the 
stabilizer states \cite{cormick06}.  Galvao has demonstrated the same result, 
and used it to support the conjecture that negativity of the discrete 
Wigner function is a necessary condition for the exponential speedup of 
quantum computation with pure states \cite{galvao05}.

\newpage

\subsection{Non-zero Temperature}
\begin{figure}[h]
\caption{\textbf{(a)} 
The time dependence (at constant bath temperature) of the negativity 
of the Wigner function for Fock states evolving in a bath of mean 
photon number $N=0.06$. Each curve is a snapshot at different values of 
time of negativity as a function of eigennumber $n$.
\textbf{(b)} 
The temperature dependence (at constant time) of negativity of the 
Wigner function as a function of $n$. Each curve is a snapshot at different 
value of the bath temperature, of the negativity as a function of
eigennumber $n$, all taken at the same unitless time $\gamma t=.15$.}
\subfigure[]{
\includegraphics[width=5in]{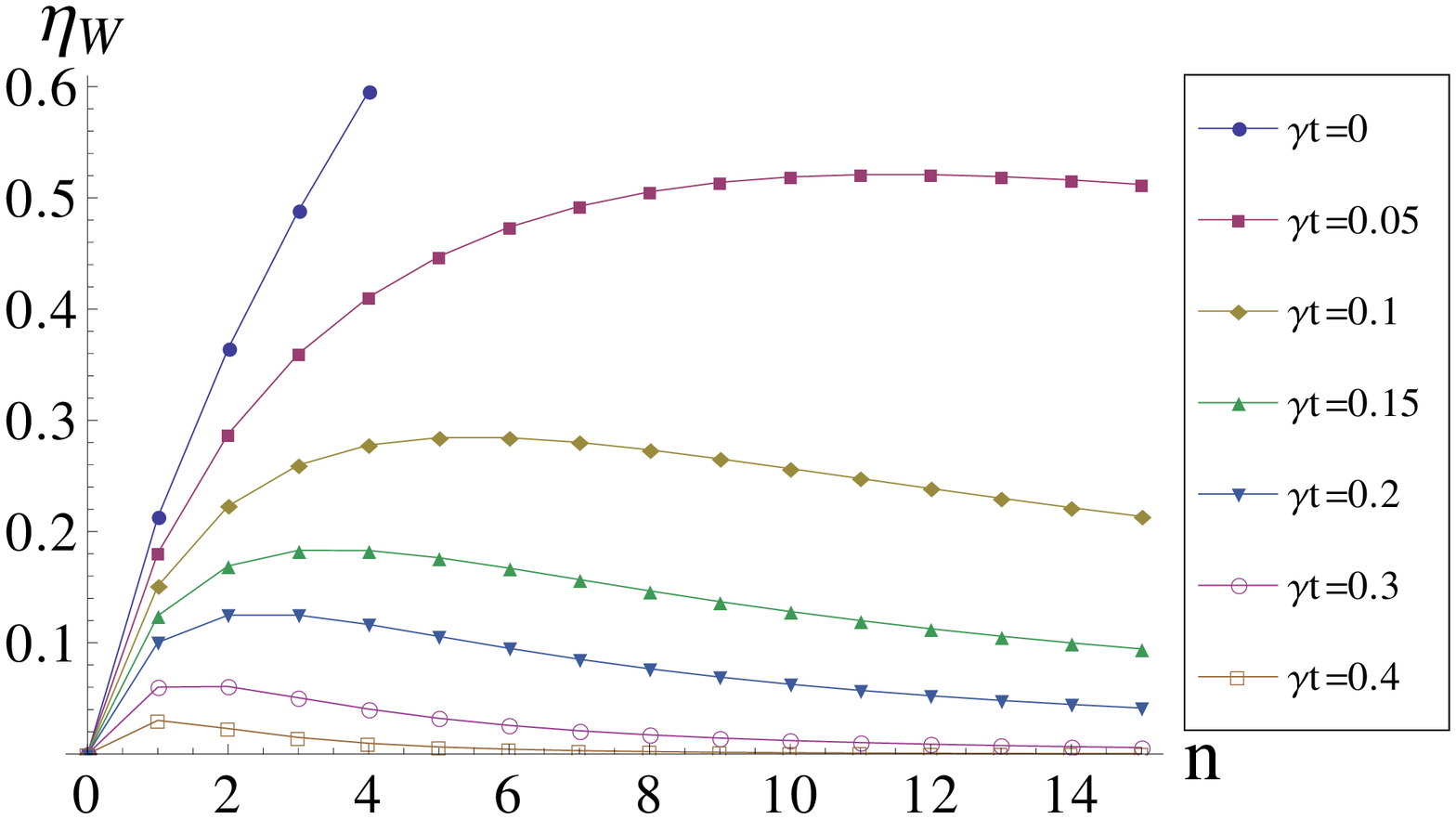}
\label{fig:negativityN0.06}}

\subfigure[]{
\includegraphics[width=5in]{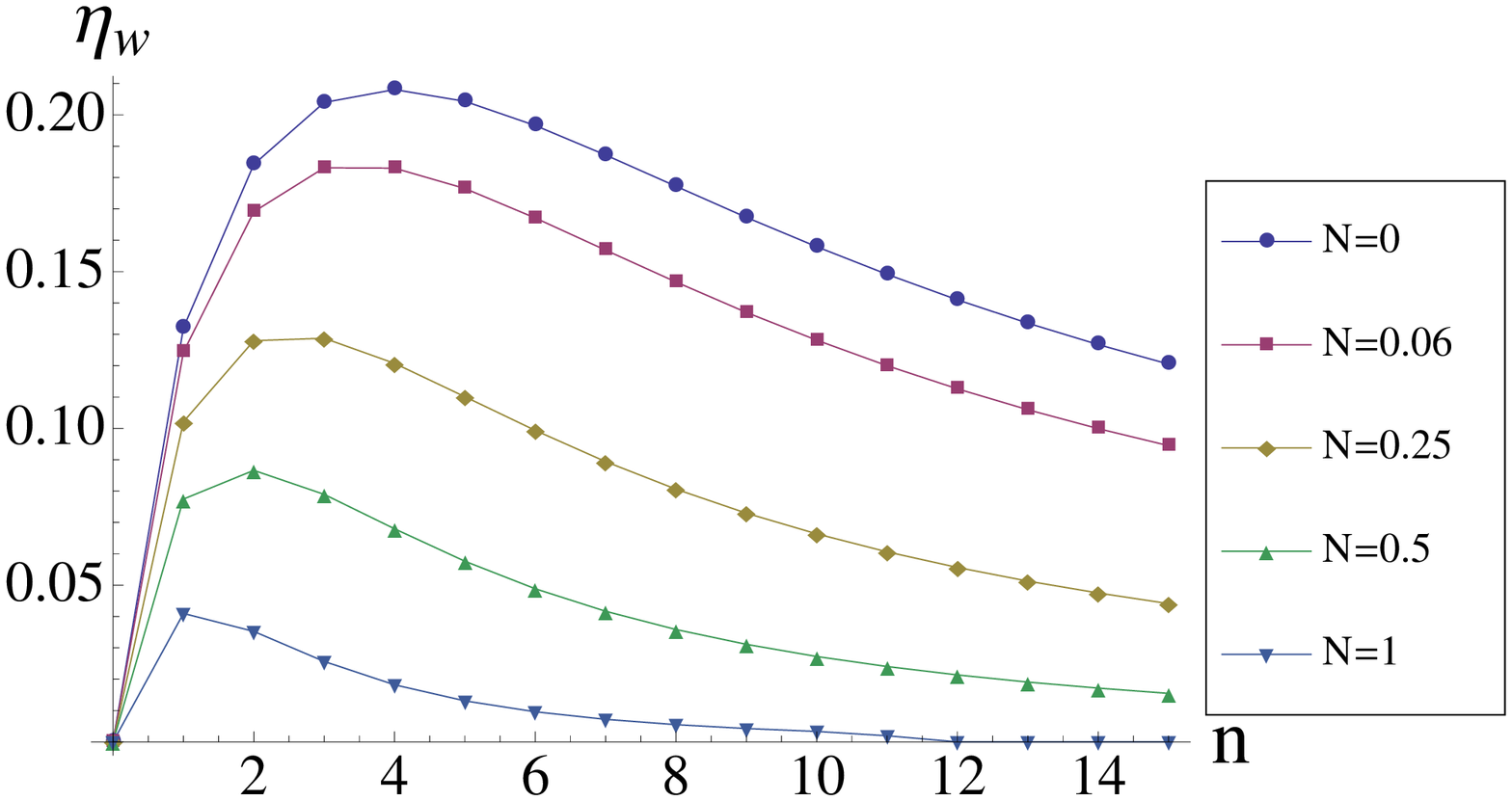}
\label{fig:negativitytau0.15}}
\label{fig:negativity}
\end{figure}

In Fig.~\ref{fig:negativityN0.06} we show negativity versus Fock state
for different values of $\gamma t$ at finite temperature ($N=0.06$).  
Notice that at $\gamma t=0$ the negativity increases monotonically in $n$, but 
that at any nonzero time there is a peak 
in negativity, again confirming our argument that the monotonicity of 
negativity at $\gamma t = 0$ is singular and unphysical.

This is reinforced by Fig.~\ref{fig:negativitytau0.15}, which shows evolution 
of the lowest 16 harmonic oscillator eigenstates to $\gamma t=0.15$ for 
different bath temperatures.  Here, as we already know from Fig.~\ref{fig:etawvsn} 
even the $N=0$ curve has a peak, in contrast with the $t=0$ peak in 
Figs.~\ref{fig:etawvsn} and \ref{fig:negativityN0.06}. We see that temperature has 
a somewhat similar effect on non-classicality as time in that higher 
temperature corresponds to greater decoherence, just as greater values of $\gamma t$
correspond to greater coherence.  On the other hand, the mere fact of $t>0$ 
leads to a peak, and the value of $\gamma t$ places an upper limit on 
what value of $n$ the non-classicality peak takes.

\subsection{Non-diagonal Initial Density Matrices}
\begin{figure}[h]
\caption{These plots show ``non-classicality" as a function of initial condition $n$ in a bath of mean-photon number $N=0.06$ at specific values of $\gamma t$ for states described by \textbf{(a)} $\psi_n[t=0]=\frac{1}{\sqrt{2}}\left(|n-1\rangle+|n\rangle\right)$, \\\textbf{(b)} $\psi_n[t=0]=\frac{1}{\sqrt{2}}\left(|n-1\rangle+|n+1\rangle\right)$, \textbf{(c)} $\psi_n[t=0]=\frac{1}{\sqrt{n}}\sum_{m=1}^n |m\rangle$, and \\ \textbf{(d)} $\psi_n[t=0]=\sum_{m=1}^{n-1} \frac{1}{2^{m/2}}|2m-1\rangle+\frac{1}{2^{(n-1)/2}}|2n-1\rangle$.}
  \begin{minipage}[b]{5.5in}
    \begin{flushleft}
\subfigure[]{
\includegraphics[width=2.6in]{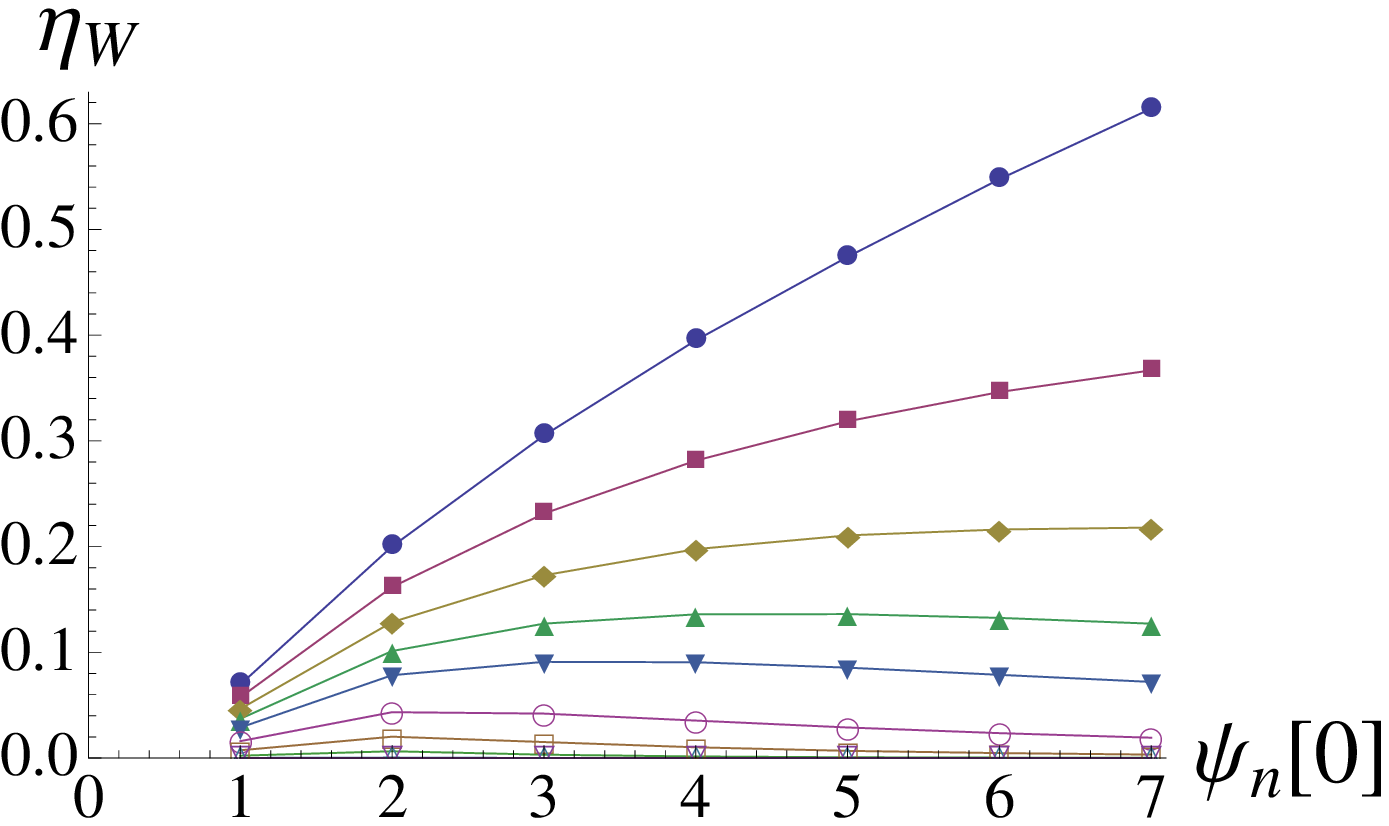}
\label{fig:consecutive}
}
\subfigure[]{
\includegraphics[width=2.6in]{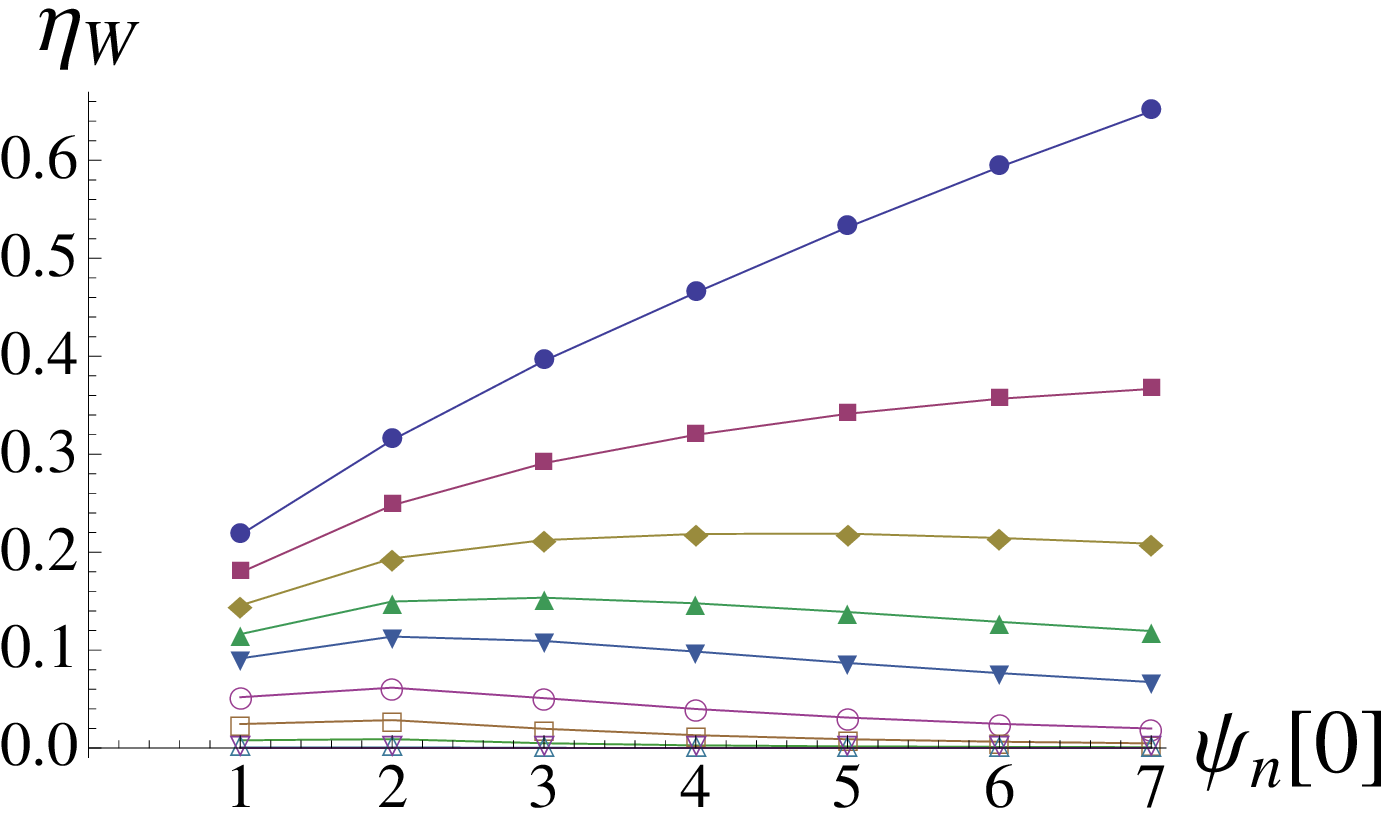}
\label{fig:nnplus2}
}

\subfigure[]{
\includegraphics[width=2.6in]{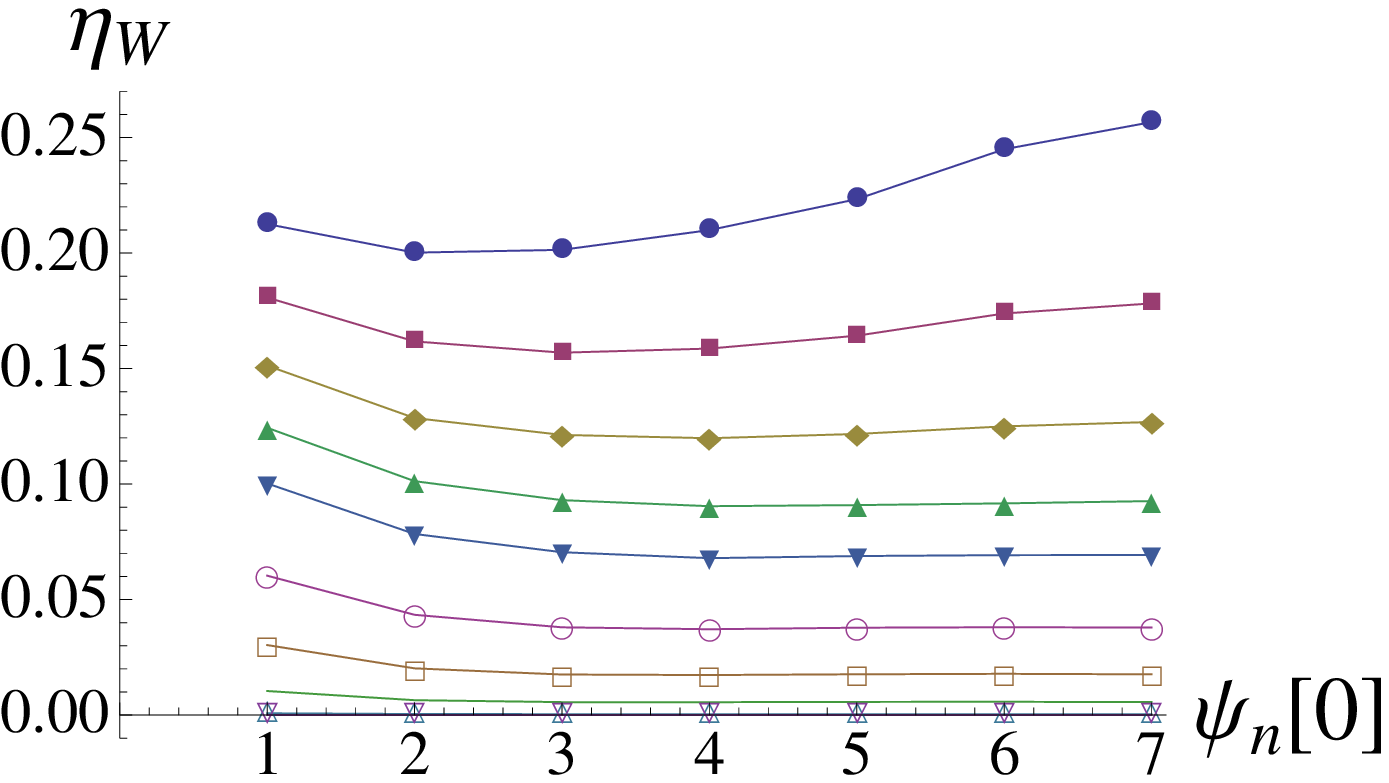}
\label{fig:equal}
}
\subfigure[]{
\includegraphics[width=2.6in]{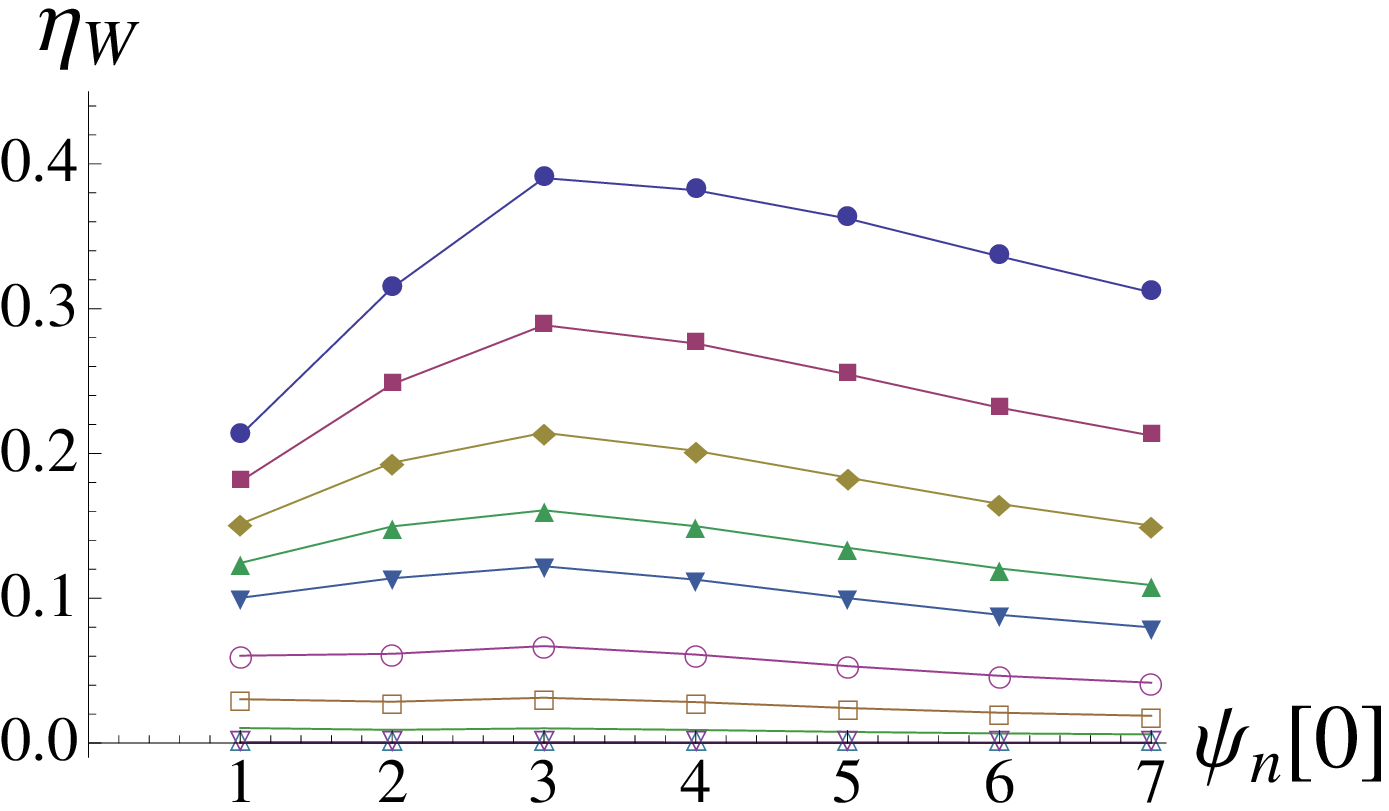}
\label{fig:coherent}
}      
    \end{flushleft}
  \end{minipage}
  \vspace{0.5cm}
  \begin{minipage}[b]{.9in}
    \begin{flushright}
      \subfigure{
        \includegraphics[width=.9in]{Legend1to15-2.eps}
        }
    \end{flushright}
  \end{minipage}
\label{fig:non-diagonal}
\end{figure}

As a final, brief foray into the rich landscape of potential states that may be explored,
we considered four other types of initial conditions (well-suited to the form of our solution), 
all evolving in a Markovian bath of 
mean-photon number $N=0.06$, which, as shown by Brune et.\ al.\, is an experimentally realizable 
temperature \cite{brune08}.  The first, displayed in Fig.\ \ref{fig:consecutive}, are of the form 
$\psi_n[t=0]=\frac{1}{\sqrt{2}}\left(|n-1\rangle+|n\rangle\right)$, and represent the simplest 
form of qubit states.  The second, displayed in Fig.\ \ref{fig:nnplus2}, are of the form 
$\psi_n[t=0]=\frac{1}{\sqrt{2}}\left(|n-1\rangle+|n+1\rangle\right)$, and represent less 
conventional qubit states.  The third, displayed in Fig.\ \ref{fig:equal}, are of the form 
$\psi_n[t=0]=\frac{1}{\sqrt{n}}\sum_{m=1}^n |m\rangle$. The fourth, displayed in 
Fig.\ \ref{fig:coherent}, are of the form $\psi_n[t=0]=\sum_{m=1}^{n-1} 
\frac{1}{2^{m/2}}|2m-1\rangle+\frac{1}{2^{(n-1)/2}}|2n-1\rangle$.  This last two are brief 
forays into the large expanse of possible qudits.

These results clearly show that in general ``non-classicality" is not as straight-forward 
for general states as for Fock states might at first suggest.  However, there are elements of 
similarity to the Fock states.  Particularly in the case of Figs.\ \ref{fig:consecutive} and 
\ref{fig:nnplus2}, we see a monotonic rise in ``non-classicality" at $t=0$ and a 
``non-classicality" peak at all other times (until all states have transitioned to classical).  
The major difference lies in the fact that the 
$\frac{1}{\sqrt{2}}|0\rangle+\frac{1}{\sqrt{2}}|1\rangle$ state never appears to cross 
the $\frac{1}{\sqrt{2}}|1\rangle+\frac{1}{\sqrt{2}}|2\rangle$ state; nor does the 
$\frac{1}{\sqrt{2}}|0\rangle+\frac{1}{\sqrt{2}}|2\rangle$ state appear to cross the 
$\frac{1}{\sqrt{2}}|1\rangle+\frac{1}{\sqrt{2}}|3\rangle$ state, so that the 
``non-classicality" peak appears to stop decreasing at $n=2$.  This is not entirely 
surprising, since each of these $n=1$ states are constructed partially from the ground state, 
which is by our definition a classical state.

The non-monotonicity of the $t=0$ curve is somewhat unexpected in Fig.\ \ref{fig:equal}.  
However, the time-dependent behavior is fairly similar to the Fock state case, in that 
ultimately the ``non-classicality" peak occurs at $n=1$ (the state corresponding to the 
initial condition $|\psi\rangle=|1\rangle$).  In contrast, the non-monotonicity of the 
$t=0$ curve in Fig.\ \ref{fig:coherent} is not so surprising.  It makes sense that at 
first as the states become more complex, the ``non-classicality" increases.  However
, ultimately these states are approaching an infinite superposition of eigenstates, rather 
similar to a coherent state, and therefore would be expected to approach
a classical state.  Here again the ``non-classicality" peak behaves somewhat unexpectedly.  
Though it eventually comes to rest at $n=1$, it skips $n=2$ entirely.

These last two may have important implications for understanding what types of states 
will be useful for qudit encoding, especially since negativity is believed to be a 
necessary condition for quantum computation speedups \cite{galvao05}
although it is the negativity of multiple entangled qudits that is
specifically important. However, it may be useful to understand the behavior of 
a single qudit first.

\section{Conclusions}
We have presented above a general solution to the harmonic oscillator linearly 
coupled to a Markovian bath, and have studied the behavior of four different
measures of non-classicality using the Fock states and special finite
superposition of Fock states. By introducing a new classical basis, 
we were able to use the three relative measures to show that this system obeys the 
correspondence principle and indicates a non-monotonic transition from
quantum to classical.  We also showed that negativity gives the same result.  

The result obtained using the three relative measures relies upon the new classical 
basis we introduce: Without this new basis, the measures appear to contradict the 
correspondence principle.  We find that negativity is arguably a more versatile and 
useful measure for studying the dynamics of systems -- particularly where an appropriate 
classical basis is harder to guess.  Finally, we used negativity to explore the 
dynamics of number states decaying in a non-zero temperature bath, and
then generalized to a few finite superpositions of eigenstates; the last serves as a 
brief survey of the rich landscape of states that are still open for study.

\begin{acknowledgements}
We thank Professor Howard Wiseman for useful comments on an earlier version of this 
paper, and funding from the Howard Hughes Medical Institute through Carleton College.
\end{acknowledgements}

\bibliography{resubbib}

\appendix
\section{Deriving the Solution}
We solved the master equation for the simple harmonic oscillator linearly coupled to a Markovian bath of oscillators, given in Eq.~\ref{master}, by representing that equation in the number basis (see Eq.~\ref{numberbasis}), and then making the substitution $m=n+k$, which leads to Eq.~\ref{numberbasisk}, and is reproduced below:
\begin{equation}
\nonumber
\frac{1}{\gamma}\dot{C}_{n,n+k} = \sqrt{n+1}\sqrt{n+k+1}(N + 1)C_{n+1,n+k+1}
\end{equation}
\begin{equation}
\label{Anumberbasisk}
- ((2n+k+1)N + \frac{2n+k}{2}) C_{n,n+k} +\sqrt{n}\sqrt{n+k}N
  C_{n-1,n+k-1}.
\end{equation}

As previously stated, Eq.~\ref{Anumberbasisk} is equivalent to the much more compact vector equation $\vec{C}_k=\hat{A}[k]\vec{C}_k$, where $\hat{A}[k]$ is an infinite 2-dimensional matrix and $\vec{C}_k$ is a vector composed of the elements $C_{n,n+k}$.  Since $\hat{A}[k]$ is an infinite dimensional tri-diagonal matrix, we cannot diagonalize it analytically.  Instead we search for the eigenvalues numerically by finding the eigenvalues of truncated matrices $\hat{A}[k]_d$ of dimension $(d+1)\times(d+1)$.  By graphing the eigenvalues as d becomes large, we see a very compelling convergence to $\lambda_j=-(j+\frac{k}{2})$, where $j$ denotes the $jth$ eigenvalue and spans the nonnegative integers.

We then take this premise as our Ansatz, and proceeded to find the eigenvectors using these eigenvalues.  We compute a finite number of entries for multiple eigenvectors until a pattern emerges.  We see that the $lth$ entry of the $jth$ eigenvector is
\begin{widetext}
\begin{equation}
\label{Tgeneral}
\sqrt{\left(\begin{array}{c}k+l\\k\end{array}\right)}\left(\frac{N}{N+1}\right)^l\sum_{i=0}^l
(-1)^i\left(\begin{array}{c}l\\i\end{array}\right) N^{-i}
\left(\begin{array}{c}j\\i\end{array}\right)\frac{1}{\left(\begin{array}{c}k+i\\k\end{array}\right)},
\end{equation}
\end{widetext}
which we use as our second Ansatz (note that we have as yet not proven the first).

We then construct finite truncations of the transformation matrix $\hat{T}$ from Eq.~(\ref{Tgeneral}) and invert them to search for the analytical form of $\hat{T}^{-1}$.  Again a pattern emerges, suggesting that the $lth,jth$ entry of $\hat{T}^{-1}$ is
\begin{widetext}
\begin{equation}
\label{Tinv}
\sqrt{\left(\begin{array}{c}k+j\\k\end{array}\right)}\left(\begin{array}{c}k+l\\k\end{array}\right)\frac{N^l}{(N+1)^{l+k+1}}\sum_{\nu=0}^l
(-1)^\nu\left(\begin{array}{c}l\\\nu\end{array}\right) N^{-\nu}
\left(\begin{array}{c}j\\\nu\end{array}\right)\frac{1}{\left(\begin{array}{c}k+\nu\\k\end{array}\right)}.
\end{equation}.
\end{widetext}
We prove that this is the inverse transformation matrix in Appendix B by showing that Eq.~(\ref{Tgeneral}) multiplied by Eq.~(\ref{Tinv}) yields the Kronecker delta.

Using the eigenvalues from our first Ansatz, we construct the diagonal matrix $\hat{D}=\hat{T}^{-1}\hat{A}[k]\hat{T}$, which is of course comprised of the eigenvalues shown along the diagonal.  In diagonal space, we have the differential equation $\frac{1}{\gamma}\frac{d}{dt}\vec{x}_k=\hat{D}\vec{x}$, where $\vec{x}_k=\hat{T}\vec{C}_k$.  The solution to this set of differential equations is trivial:
\begin{equation}
x_{n,n+k}=a_n e^{\gamma \lambda_n t}
\end{equation}
and can be expressed in vector notation as $\vec{x}_k=e^{-\gamma kt/2}\hat{E}\vec{a}$, where
\begin{equation}
\hat{E}=\left(\begin{array}{cccccc}\ddots&\,&\,&\,&\,&\vdots\\\,&e^{-n\gamma t}&0&\dots&\,&0\\\,&\,&\ddots&&\,&\,\vdots\\\,&\ddots&0&e^{-2\gamma t}&0&0\\\,&\dots&0&0&e^{-\gamma t}&0\\\,&\dots&0&0&0&1\end{array}\right)
\end{equation}
and $\vec{a}$ is composed of the unknown constants $a_n$.

Transforming back into the original space, we have $\vec{C}_k[t]=\hat{T}e^{-\gamma kt/2}\hat{E}\vec{a}$, or
\begin{widetext}
\begin{equation}
\label{AgeneralC}
C_{n,n+k}=\sum_{j=0}^\infty
a_j\sqrt{\left(\begin{array}{c}k+n\\k\end{array}\right)}\left(\frac{N}{N+1}\right)^n\sum_{i=0}^n
(-1)^i\left(\begin{array}{c}n\\i\end{array}\right) N^{-i}
\left(\begin{array}{c}j\\i\end{array}\right)\frac{1}{\left(\begin{array}{c}k+i\\k\end{array}\right)}e^{-\gamma(j+\frac{k}{2})t}.
\end{equation}
\end{widetext}
The constants $a_n$ depend upon the initial conditions, and can be solved by setting $t=0$: $\vec{C}_k[0]=\hat{T}\vec{a}$, or $\vec{a}=\hat{T}^{-1}\vec{C}_k[0]$.  We now have all the tools we need to write down the final solution in vector notation: 
\begin{equation}
\label{finalvectorform}
\vec{C}_k[t]=\hat{T}e^{-\gamma kt/2}\hat{E}\hat{T}^{-1}\vec{C}_k[0].
\end{equation}  

In order to find the closed form expression for an entry of a general density matrix $C_{n,n+k}$, we define the initial conditions to be 
\begin{equation}
C_{n,n+k}[0]=\left\{\begin{array}{c}1,n=n_0\\0,n\neq n_0\end{array}\right. ,
\end{equation}
which gives rise to Eq.~(\ref{partialsolution}) upon substituting them into Eq.~(\ref{finalvectorform}).  Since these initial condition vectors represent an orthonormal set, Eq.~(\ref{partialsolution}) can be used to find the solution to any initial condition.  This idea is represented mathematically in Eq.~(\ref{rhosimp}).

However, this solution rests on two Ans\"atze: that the eigenvalues of $\hat{A}[k]$ are $\lambda_j=-(j+\frac{k}{2})$ and that the eigenvectors are given by Eq.~(\ref{Tgeneral}).  These Ans\"atze are proven by substituting Eq.~(\ref{generalC}) into Eq.~(\ref{numberbasisk}), and showing that the solution we found does in fact solve the original differential equation (see Appendix C).  Since there are sufficient $a_n$ to span the entire solution space, the uniqueness and existence theorem guarantees that our solution is correct, and by extension that our Ans\"atze were correct.

\section{Inversion Proof}
In order to prove that the matrix, here denoted $\hat{B}$, given by equation (\ref{Tinv}) and reproduced below
\begin{equation}
\sqrt{\left(\begin{array}{c}k+j\\k\end{array}\right)}\left(\begin{array}{c}k+l\\k\end{array}\right)\frac{N^l}{(N+1)^{l+k+1}}\sum_{\nu=0}^l (-1)^\nu\left(\begin{array}{c}l\\\nu\end{array}\right) N^{-\nu} \left(\begin{array}{c}j\\\nu\end{array}\right)\frac{1}{\left(\begin{array}{c}k+\nu\\k\end{array}\right)}
\end{equation}
is the inverse of $\hat{T}$, given by equation (\ref{Tgeneral}) and also reproduced below
\begin{equation}
\sqrt{\left(\begin{array}{c}k+l\\k\end{array}\right)}\left(\frac{N}{N+1}\right)^l\sum_{i=0}^l (-1)^i\left(\begin{array}{c}l\\i\end{array}\right) N^{-i} \left(\begin{array}{c}j\\i\end{array}\right)\frac{1}{\left(\begin{array}{c}k+i\\k\end{array}\right)}
\end{equation}

we will show that the $ath$ row of $\hat{T}$ multiplied by the $bth$ column of $\hat{B}$ is the Kronecker delta:
\begin{equation}
\nonumber
\sqrt{\left(\begin{array}{c}k+a\\k\end{array}\right)\left(\begin{array}{c}k+b\\k\end{array}\right)}\frac{N^a}{(N+1)^{a+k+1}}\sum_{i=0}^a \sum_{\nu=0}^b (-1)^{i+\nu}N^{-i-\nu}\left(\begin{array}{c}a\\i\end{array}\right)  \left(\begin{array}{c}b\\\nu\end{array}\right)\frac{1}{\left(\begin{array}{c}k+\nu\\k\end{array}\right)\left(\begin{array}{c}k+i\\k\end{array}\right)}
\end{equation}
\begin{equation}
\label{TB}
\times \left[\sum_{m=0}^\infty \left(\begin{array}{c}m\\i\end{array}\right)\left(\begin{array}{c}m\\\nu\end{array}\right)\left(\begin{array}{c}m+k\\k\end{array}\right)\left(\frac{N}{N+1}\right)^m\right] = \delta_{a,b}.
\end{equation}

Dealing with the infinite sum in square brackets (letting $\left(\frac{N}{N+1}\right)=r$), it is easy to show that
\begin{equation}
\label{modgeo2}
\sum_{m=0}^\infty \left(\begin{array}{c}m\\i\end{array}\right)\left(\begin{array}{c}m\\\nu\end{array}\right)\left(\begin{array}{c}m+k\\k\end{array}\right)r^m =
\frac{1}{k!}\frac{d^k}{dr^k}r^k \left\{\frac{r^l}{l!}\frac{d^l}{dr^l}\left(\frac{r^i}{(1-r)^{i+1}}\right)\right\}.
\end{equation}

We first simplify the term in curly braces.  Using the fact that:
\begin{equation}
\frac{d^l}{dr^l}(f[r]g[r])
=
\sum_{\nu=0}^l \left(\begin{array}{c}l\\\nu\end{array}\right) f^{(l-\nu)}[r]g^{(\nu)}[r],
\end{equation}
and setting $f[r]=(1-r)^{-i-1}$ and $g[r]=r^i$ we have:
\begin{equation}
\frac{r^l}{l!}\frac{d^l}{dr^l}\left(\frac{r^i}{(1-r)^{i+1}}\right)
=
\frac{r^l}{l!}\sum_{\nu=0}^l \left(\begin{array}{c}l\\\nu\end{array}\right)\frac{(i+l-\nu)!}{i!}(1-r)^{-i-l-1+\nu}\frac{i!}{(i-l)!}r^{i-\nu}.
\end{equation}
which simplifies to
\begin{equation}
\frac{r^{i+l}}{(1-r)^{i+l+1}}\sum_{\nu=0}^l \left(\begin{array}{c}l\\\nu\end{array}\right) \left(\begin{array}{c}i+l-\nu\\l\end{array}\right)\left(\frac{1-r}{r}\right)^\nu.
\end{equation}

By the binomial expansion this becomes:
\begin{equation}
\frac{r^{i+l}}{(1-r)^{i+l+1}}\sum_{\nu=0}^l\sum_{\mu=0}^\nu \left(\begin{array}{c}l\\\nu\end{array}\right) \left(\begin{array}{c}i+l-\nu\\l\end{array}\right)\left(\begin{array}{c}\nu\\\mu\end{array}\right)(-1)^{\mu+\nu}r^{-\mu}.
\end{equation}
Reordering the sum and grouping by $r^{-\mu}$ we have:
\begin{equation}
\frac{r^{i+l}}{(1-r)^{i+l+1}}\sum_{\mu=0}^l r^{-\mu}\sum_{\nu=\mu}^l \left(\begin{array}{c}l\\\nu\end{array}\right) \left(\begin{array}{c}i+l-\nu\\l\end{array}\right)\left(\begin{array}{c}\nu\\\mu\end{array}\right)(-1)^{\mu+\nu}.
\end{equation}

The coefficients of $r^{-\mu}$ can be simplified by summing over $\nu$.  We first reindex the sum by setting $\nu=\mu+k$:
\begin{equation}
\sum_{\nu=\mu}^l \left(\begin{array}{c}l\\\nu\end{array}\right) \left(\begin{array}{c}i+l-\nu\\l\end{array}\right)\left(\begin{array}{c}\nu\\\mu\end{array}\right)(-1)^{\mu+\nu}
=
\sum_{k=0}^{l-\mu} \left(\begin{array}{c}l\\\mu+k\end{array}\right) \left(\begin{array}{c}i+l-\mu-k\\l\end{array}\right)\left(\begin{array}{c}\mu+k\\\mu\end{array}\right)(-1)^{\mu+\nu}.
\end{equation}
The right-hand sum can be rewritten as a multiple of a hypergeometric series:
\begin{equation}
\frac{(i+l-\mu)!}{(l-\mu)!(i-\mu)!\mu!} \,_2F_1[-(l-\mu), \mu-i; \mu-i-l; 1],
\end{equation}
allowing us to use the Chu-Vandermonde identity, which states that
\begin{equation}
_2F_1[-n, \beta; \gamma; 1]=\frac{(\gamma-\beta)_n}{(\gamma)_n},
\end{equation}
where $(x)_n=x(x+1)...(x+n-1); n>0$ is the Pochanner symbol.  Substituting this solution for the hypergeometric series yields
\begin{equation}
\sum_{k=0}^{l-\mu} \left(\begin{array}{c}l\\\mu+k\end{array}\right) \left(\begin{array}{c}i+l-\mu-k\\l\end{array}\right)\left(\begin{array}{c}\mu+k\\\mu\end{array}\right)(-1)^{\mu+\nu}
=
\frac{(i+l-\mu)!}{(l-\mu)!(i-\mu)!\mu!}\frac{(-l)_{l-\mu}}{(\mu-i-l)_{l-\mu}},
\end{equation}
which can be rewritten as
\begin{equation}
\frac{(i+l-\mu)!}{(l-\mu)!(i-\mu)!\mu!}\frac{l!i!}{\mu!(i+l-\mu)!}=\frac{i!l!}{(l-\mu)!\mu!(i-\mu)!\mu!}.
\end{equation}
Therefore
\begin{equation}
\sum_{\nu=\mu}^l \left(\begin{array}{c}l\\\nu\end{array}\right) \left(\begin{array}{c}i+l-\nu\\l\end{array}\right)\left(\begin{array}{c}\nu\\\mu\end{array}\right)(-1)^{\mu+\nu}
=
\left(\begin{array}{c}l\\\nu\end{array}\right)\left(\begin{array}{c}i\\\nu\end{array}\right)
\end{equation}
and
\begin{equation}
\label{geosolved}
\sum_{m=0}^\infty \left(\begin{array}{c}m\\l\end{array}\right)\left(\begin{array}{c}m\\i\end{array}\right)r^m
=
\frac{r^{i+l}}{(1-r)^{i+l+1}}\sum_{\nu=0}^l\left(\begin{array}{c}l\\\nu\end{array}\right)\left(\begin{array}{c}i\\\nu\end{array}\right) r^{-\nu}
\end{equation}

Returning now to Eq.~(\ref{modgeo2}), we have
\begin{equation}
\frac{1}{k!}\frac{d^k}{dr^k}\left[ r^k \frac{r^{i+\nu}}{(1-r)^{i+\nu+1}}\sum_{j=0}^\nu \left(\begin{array}{c}i\\j\end{array}\right)\left(\begin{array}{c}\nu\\j\end{array}\right)r^{-j}\right],
\end{equation}
which can be written as:
\begin{equation}
\frac{1}{k!} \sum_{j=0}^\nu \left(\begin{array}{c}i\\j\end{array}\right)\left(\begin{array}{c}\nu\\j\end{array}\right)\sum_{\mu=0}^k\left(\begin{array}{c}k\\\mu\end{array}\right)\frac{d^{k-\mu}}{dr^{k-\mu}}\left((1-r)^{-i-\nu-1}\right)\frac{d^{\mu}}{dr^{\mu}}\left(r^{i+\nu+k-j}\right).
\end{equation}

This becomes
\begin{equation}
\frac{r^{i+\nu+k}}{(1-r)^{i+\nu+k+1}}\sum_{j=0}^\nu \left(\begin{array}{c}i\\j\end{array}\right)\left(\begin{array}{c}\nu\\j\end{array}\right)r^{-j}\sum_{\mu=0}^k\left(\begin{array}{c}i+\nu+k-\mu\\i+\nu\end{array}\right)\left(\begin{array}{c}i+\nu+k-j\\\mu\end{array}\right)\left(\frac{1-r}{r}\right)^{\mu}
\end{equation}
and by the binomial expansion we have
\begin{equation}
\frac{r^{i+\nu+k}}{(1-r)^{i+\nu+k+1}}\sum_{j=0}^\nu \left(\begin{array}{c}i\\j\end{array}\right)\left(\begin{array}{c}\nu\\j\end{array}\right)r^{-j}\sum_{\mu=0}^k\left(\begin{array}{c}i+\nu+k-\mu\\i+\nu\end{array}\right)\left(\begin{array}{c}i+\nu+k-j\\\mu\end{array}\right)\sum_{w=0}^\mu \left(\begin{array}{c}\mu\\w\end{array}\right)(-1)^{\mu+w}r^{-w}.
\end{equation}

By rearranging the sums over $w$ and $\mu$ this expression can be rewritten as
\begin{equation}
\label{blah}
\frac{r^{i+\nu+k}}{(1-r)^{i+\nu+k+1}}\sum_{j=0}^\nu \left(\begin{array}{c}i\\j\end{array}\right)\left(\begin{array}{c}\nu\\j\end{array}\right)r^{-j}\sum_{w=0}^k r^{-w} \left[\sum_{\mu=w}^k\left(\begin{array}{c}i+\nu+k-\mu\\i+\nu\end{array}\right)\left(\begin{array}{c}i+\nu+k-j\\\mu\end{array}\right) \left(\begin{array}{c}\mu\\w\end{array}\right)(-1)^{\mu+w}\right].
\end{equation}

The term from (\ref{blah}) in square brackets is actually a hypergeometric series, and can be written as
\begin{equation}
\frac{(i+\nu+k-j)!(i+\nu+k-w)!}{(i+\nu)!w!(k-w)!(j+w-k)!}\,_2F_1[-(k-w),(w+j-k-i-\nu);(w-k-i-\nu);1],
\end{equation}
which allows us to apply the Chu-Vandermonde identity again to arrive at
\begin{equation}
\sum_{\mu=w}^k\left(\begin{array}{c}i+\nu+k-\mu\\i+\nu\end{array}\right)\left(\begin{array}{c}i+\nu+k-j\\\mu\end{array}\right) \left(\begin{array}{c}\mu\\w\end{array}\right)(-1)^{\mu+w}=
\left(\begin{array}{c}i+\nu+k-j\\w\end{array}\right)\left(\begin{array}{c}j\\k-w\end{array}\right).
\end{equation}

Therefore
\begin{equation}
\sum_{m=0}^\infty \left(\begin{array}{c}m\\i\end{array}\right)\left(\begin{array}{c}m\\\nu\end{array}\right)\left(\begin{array}{c}m+k\\k\end{array}\right)r^m
\end{equation}
\begin{equation}
=
\frac{r^{i+\nu+k}}{(1-r)^{i+\nu+k+1}}\sum_{j=0}^\nu \left(\begin{array}{c}i\\j\end{array}\right)\left(\begin{array}{c}\nu\\j\end{array}\right)r^{-j}\sum_{w=0}^k \left(\begin{array}{c}i+\nu+k-j\\w\end{array}\right)\left(\begin{array}{c}j\\k-w\end{array}\right) r^{-w}.
\end{equation}

Recalling that $r=\frac{N}{N+1}$, equation (\ref{TB}) can be written without the infinite sum as
\begin{equation}
\nonumber
\sqrt{\left(\begin{array}{c}k+a\\k\end{array}\right)\left(\begin{array}{c}k+b\\k\end{array}\right)}\left(\frac{N}{N+1}\right)^{a+k}\sum_{i=0}^a \sum_{\nu=0}^b \sum_{j=0}^\nu \sum_{w=0}^k (-1)^{i+\nu}\left(\begin{array}{c}a\\i\end{array}\right)  \left(\begin{array}{c}b\\\nu\end{array}\right)\frac{1}{\left(\begin{array}{c}k+\nu\\k\end{array}\right)\left(\begin{array}{c}k+i\\k\end{array}\right)}
\end{equation}
\begin{equation}
\times \left(\begin{array}{c}i\\j\end{array}\right)\left(\begin{array}{c}\nu\\j\end{array}\right) \left(\frac{N}{N+1}\right)^{-j} \left(\begin{array}{c}i+\nu+k-j\\w\end{array}\right)\left(\begin{array}{c}j\\k-w\end{array}\right) \left(\frac{N}{N+1}\right)^{-w}.
\end{equation}
Rearranging the sums over $j$ and $w$ yields:
\begin{equation}
\nonumber
\sqrt{\left(\begin{array}{c}k+a\\k\end{array}\right)\left(\begin{array}{c}k+b\\k\end{array}\right)}\left(\frac{N}{N+1}\right)^{a+k}\sum_{i=0}^a \sum_{\nu=0}^b \sum_{j=0}^\nu (-1)^{i+\nu}\left(\begin{array}{c}a\\i\end{array}\right)  \left(\begin{array}{c}b\\\nu\end{array}\right)\frac{1}{\left(\begin{array}{c}k+\nu\\k\end{array}\right)\left(\begin{array}{c}k+i\\k\end{array}\right)}
\end{equation}
\begin{equation}
\label{blah2}
\times  \left(\frac{N+1}{N}\right)^{j+k}\left[\sum_{w=0}^{k} \left(\begin{array}{c}i\\w+j\end{array}\right)\left(\begin{array}{c}\nu\\w+j\end{array}\right)\left(\begin{array}{c}i+\nu+k-j-w\\k-w\end{array}\right)\left(\begin{array}{c}w+j\\j\end{array}\right)\right].
\end{equation}
The term in square brackets in Eq.~(\ref{blah2}) is a $_3F_2$ hypergeometric series:
\begin{equation}
\nonumber
\sum_{w=0}^{k} \left(\begin{array}{c}i\\w+j\end{array}\right)\left(\begin{array}{c}\nu\\w+j\end{array}\right)\left(\begin{array}{c}i+\nu+k-j-w\\k-w\end{array}\right)\left(\begin{array}{c}w+j\\j\end{array}\right)
\end{equation}
\begin{equation}
=
\left(\begin{array}{c}i\\j\end{array}\right)\left(\begin{array}{c}\nu\\j\end{array}\right)\left(\begin{array}{c}i+\nu+k-j\\k\end{array}\right)\, _3F_2[(-k),(j-i),(j-\nu);(j+1),(j-i-\nu-k);1]
\end{equation}
which can be simplified to
\begin{equation}
\left(\begin{array}{c}i\\j\end{array}\right)\left(\begin{array}{c}\nu\\j\end{array}\right)\left(\begin{array}{c}i+\nu+k-j\\k\end{array}\right)
\frac{\left(\begin{array}{c}i+k\\k\end{array}\right)\left(\begin{array}{c}\nu+k\\k\end{array}\right)}{\left(\begin{array}{c}j+k\\k\end{array}\right)\left(\begin{array}{c}i+\nu+k-j\\k\end{array}\right)}
\end{equation}
by use of the Pfaff-Saalschutz identity, which states that
\begin{equation}
_3F_2[-n,a,b;c,1+a+b-c-n;1]=\frac{(c-a)_n(c-b)_n}{(c)_n(c-a-b)_n}.
\end{equation}

Thus equation (\ref{TB}) reduces to
\begin{equation}
\sqrt{\left(\begin{array}{c}k+a\\k\end{array}\right)\left(\begin{array}{c}k+b\\k\end{array}\right)}\left(\frac{N}{N+1}\right)^{a}\sum_{i=0}^a \sum_{\nu=0}^b \sum_{j=0}^{min(i,\nu)} (-1)^{i+\nu}\left(\begin{array}{c}a\\i\end{array}\right)  \left(\begin{array}{c}b\\\nu\end{array}\right)\left(\frac{N+1}{N}\right)^{j}\frac{\left(\begin{array}{c}i\\j\end{array}\right)\left(\begin{array}{c}\nu\\j\end{array}\right)}{\left(\begin{array}{c}j+k\\k\end{array}\right)}.
\end{equation}

Rearranging the terms for clarity, we have:
\begin{equation}
\sqrt{\left(\begin{array}{c}k+a\\k\end{array}\right)\left(\begin{array}{c}k+b\\k\end{array}\right)}\left(\frac{N}{N+1}\right)^{a}\sum_{i=0}^a \sum_{\nu=0}^b \sum_{j=0}^{min(i,\nu)} (-1)^{i+\nu}\left(\begin{array}{c}a\\i\end{array}\right) \left(\begin{array}{c}i\\j\end{array}\right) \left(\begin{array}{c}b\\\nu\end{array}\right) \left(\begin{array}{c}\nu\\j\end{array}\right)\frac{\left(\frac{N+1}{N}\right)^{j}}{\left(\begin{array}{c}j+k\\k\end{array}\right)}.
\end{equation}
Calling the combinatorial identity for the kronecker delta 
\begin{equation}
\label{kroneckersimple}
\sum_{m=0}^i
(-1)^{m+j}\left(\begin{array}{c}i\\m\end{array}\right)\left(\begin{array}{c}m\\j\end{array}\right)=\delta_{i,j}
\end{equation}
twice we have
\begin{equation}
\sqrt{\left(\begin{array}{c}k+a\\k\end{array}\right)\left(\begin{array}{c}k+b\\k\end{array}\right)}\left(\frac{N}{N+1}\right)^{a}\sum_{j=0}^{min(a,b)} \\\delta_{a,j} \delta_{b,j} \frac{\left(\frac{N+1}{N}\right)^{j}}{\left(\begin{array}{c}j+k\\k\end{array}\right)}
=
\delta_{a,b}.
\end{equation}
Therefore $\hat{T}\hat{B}=\hat{I}$, and we have proven that $\hat{B}=\hat{T}^{-1}$.

\section{Uniqueness and Existence Proof}
We prove that equation (\ref{generalC}), reproduced below
\begin{equation}
\label{generalCA}
C_{n,n+k}=\sum_{j=0}^\infty a_j\sqrt{\left(\begin{array}{c}k+n\\k\end{array}\right)}\left(\frac{N}{N+1}\right)^n\sum_{i=0}^n (-1)^i\left(\begin{array}{c}n\\i\end{array}\right) N^{-i} \left(\begin{array}{c}j\\i\end{array}\right)\frac{1}{\left(\begin{array}{c}k+i\\k\end{array}\right)}e^{-\gamma(j+\frac{k}{2})t}
\end{equation}
is the solution to equation (\ref{numberbasisk}), reproduced below
\begin{equation}
\nonumber
\frac{1}{\gamma}\dot{C}_{n,n+k} = \sqrt{n+1}\sqrt{n+k+1}(N + 1)C_{n+1,n+k+1}
\end{equation}
\begin{equation}
\label{numberbasiskA}
- ((2n+k+1)N + \frac{2n+k}{2}) C_{n,n+k} +\sqrt{n}\sqrt{n+k}N C_{n-1,n+k-1}
\end{equation}
by substituting equation (\ref{generalCA}) into equation (\ref{numberbasiskA}) and showing that both sides are equal.

We start by breaking the righthand side of equation (\ref{numberbasiskA}) into three parts:
\begin{equation}
\nonumber
\left(\sqrt{n+1}\sqrt{n+k+1}(N + 1)C_{n+1,n+k+1}-(n+k+1)N  C_{n,n+k}\right)
\end{equation}
\begin{equation}
\label{righthand}
+\left\{-n(N+1) C_{n,n+k} + \sqrt{n}\sqrt{n+k}N C_{n-1,n+k-1}\right\} -
\left[\frac{k}{2} C_{n,n+k}\right].
\end{equation}
Comparing this to the lefthand side
\begin{equation}
\label{lefthand}
\sum_{j=0}^\infty a_j\sqrt{\left(\begin{array}{c}k+n\\k\end{array}\right)}\left(\frac{N}{N+1}\right)^n\sum_{i=0}^n (-1)^i\left(\begin{array}{c}n\\i\end{array}\right) N^{-i} \left(\begin{array}{c}j\\i\end{array}\right)\frac{1}{\left(\begin{array}{c}k+i\\k\end{array}\right)}(-j-\frac{k}{2})e^{-\gamma(j+\frac{k}{2})t}
\end{equation}
we see that the term in square brackets in equation (\ref{righthand}) also occurs on the lefthand side.

Next we consider the term in parentheses from equation (\ref{righthand}).  
\begin{equation}
\nonumber
\sqrt{n+1}\sqrt{n+k+1}(N + 1)C_{n+1,n+k+1} =
\end{equation}
\begin{equation}
\label{parentheses1}
(n+k+1)\frac{N^{n+1}}{(N+1)^n}\sqrt{\left(\begin{array}{c}k+n\\k\end{array}\right)}\sum_{j=0}^\infty\sum_{i=0}^{n+1} (-1)^i\left(\begin{array}{c}n+1\\i\end{array}\right) N^{-i} \left(\begin{array}{c}j\\i\end{array}\right)\frac{1}{\left(\begin{array}{c}k+i\\k\end{array}\right)}e^{-\gamma(j+\frac{k}{2})t}
\end{equation}
and
\begin{equation}
\nonumber
(n+k+1)N  C_{n,n+k}
=
\end{equation}
\begin{equation}
\label{parentheses2}
(n+k+1)\frac{N^{n+1}}{(N+1)^n}\sqrt{\left(\begin{array}{c}k+n\\k\end{array}\right)}\sum_{j=0}^\infty\sum_{i=0}^{n} (-1)^i\left(\begin{array}{c}n\\i\end{array}\right) N^{-i} \left(\begin{array}{c}j\\i\end{array}\right)\frac{1}{\left(\begin{array}{c}k+i\\k\end{array}\right)}e^{-\gamma(j+\frac{k}{2})t}
\end{equation}

Using the binomial identity
\begin{equation}
\label{binomialaddition}
\left(\begin{array}{c}n\\k\end{array}\right)=\left(\begin{array}{c}n-1\\k\end{array}\right)+\left(\begin{array}{c}n-1\\k-1\end{array}\right)
\end{equation}
we subtract equation (\ref{parentheses2}) from equation (\ref{parentheses1}) and reindex the sum by $i\rightarrow i+1$ to get
\begin{equation}
\nonumber
\sqrt{n+1}\sqrt{n+k+1}(N + 1)C_{n+1,n+k+1}-(n+k+1)N  C_{n,n+k}=
\end{equation}
\begin{equation}
\label{parenthesestotal}
-(n+k+1)\left(\frac{N}{N+1}\right)^n
\sqrt{\left(\begin{array}{c}k+n\\k\end{array}\right)}\sum_{j=0}^\infty\sum_{i=0}^{n}
(-1)^i\left(\begin{array}{c}n\\i\end{array}\right) N^{-i}
\left(\begin{array}{c}j\\i+1\end{array}\right)\frac{1}{\left(\begin{array}{c}k+1+i\\k\end{array}\right)}a_je^{-\gamma(j+\frac{k}{2})t}.
\end{equation}
Now we consider the term from equation (\ref{righthand}) in curly brackets
\begin{equation}
\nonumber
-n(N+1) C_{n,n+k} =
\end{equation}
\begin{equation}
\label{curly1}
-n\frac{N^{n}}{(N+1)^{n-1}}\sqrt{\left(\begin{array}{c}k+n\\k\end{array}\right)}\sum_{j=0}^\infty\sum_{i=0}^{n} (-1)^i\left(\begin{array}{c}n\\i\end{array}\right) N^{-i} \left(\begin{array}{c}j\\i\end{array}\right)\frac{1}{\left(\begin{array}{c}k+i\\k\end{array}\right)}a_je^{-\gamma(j+\frac{k}{2})t}
\end{equation}
and
\begin{equation}
\nonumber
\sqrt{n}\sqrt{n+k}N C_{n-1,n+k-1}
=
\end{equation}
\begin{equation}
\label{curly2}
-n\frac{N^{n}}{(N+1)^{n-1}}\sqrt{\left(\begin{array}{c}k+n\\k\end{array}\right)}\sum_{j=0}^\infty\sum_{i=0}^{n-1} (-1)^i\left(\begin{array}{c}n-1\\i\end{array}\right) N^{-i} \left(\begin{array}{c}j\\i\end{array}\right)\frac{1}{\left(\begin{array}{c}k+i\\k\end{array}\right)}a_je^{-\gamma(j+\frac{k}{2})t}.
\end{equation}
We subtract Eq.~(\ref{curly1}) from Eq.~(\ref{curly2}), once again
using Eq.~(\ref{binomialaddition}) and reindexing the sum by $i\rightarrow i+1$, to get
\begin{equation}
\nonumber
-n(N+1) C_{n,n+k} + \sqrt{n}\sqrt{n+k}N C_{n-1,n+k-1}=
\end{equation}
\begin{equation}
\label{curlytotal}
n\left(\frac{N}{N+1}\right)^{n-1}
\sqrt{\left(\begin{array}{c}k+n\\k\end{array}\right)}\sum_{j=0}^\infty\sum_{i=0}^{n-1}
(-1)^i\left(\begin{array}{c}n-1\\i\end{array}\right) N^{-i}
\left(\begin{array}{c}j\\i+1\end{array}\right)\frac{1}{\left(\begin{array}{c}k+1+i\\k\end{array}\right)}a_je^{-\gamma(j+\frac{k}{2})t}.
\end{equation}
If we subtract the lefthand side of Eq.~(\ref{numberbasiskA}) from both
sides, the term in Eq.~(\ref{righthand}) in square brackets cancels, and the remaining terms are
\begin{equation}
\label{A12}
0=\left(\frac{N}{N+1}\right)^n\sqrt{\left(\begin{array}{c}k+n\\k\end{array}\right)}\sum_{j=0}^\infty a_je^{-\gamma(j+\frac{k}{2})t}\sum_{i=0}^{n}(-\frac{1}{N})^i\bigg[-(n+k+1)\frac{\left(\begin{array}{c}n\\i\end{array}\right)\left(\begin{array}{c}j\\i+1\end{array}\right)}{\left(\begin{array}{c}k+1+i\\k\end{array}\right)}\nonumber
\end{equation}
\begin{equation}
+n(1+N^{-1})\frac{\left(\begin{array}{c}n-1\\i\end{array}\right)\left(\begin{array}{c}j\\i+1\end{array}\right)}{\left(\begin{array}{c}k+1+i\\k\end{array}\right)}+j\frac{\left(\begin{array}{c}n\\i\end{array}\right)\left(\begin{array}{c}j\\i\end{array}\right)}{\left(\begin{array}{c}k+i\\k\end{array}\right)}\bigg].
\end{equation}
From here we examine the series over $i$.  Recognizing that the constant term is zero, the series can be rearranged to be
\begin{equation}
\sum_{i=1}^{n}(-\frac{1}{N})^i\bigg[-(n+k+1)\frac{\left(\begin{array}{c}n\\i\end{array}\right)\left(\begin{array}{c}j\\i+1\end{array}\right)}{\left(\begin{array}{c}k+1+i\\k\end{array}\right)}
+n\frac{\left(\begin{array}{c}n-1\\i\end{array}\right)\left(\begin{array}{c}j\\i+1\end{array}\right)}{\left(\begin{array}{c}k+1+i\\k\end{array}\right)}-n\frac{\left(\begin{array}{c}n-1\\i-1\end{array}\right)\left(\begin{array}{c}j\\i\end{array}\right)}{\left(\begin{array}{c}k+i\\k\end{array}\right)}+j\frac{\left(\begin{array}{c}n\\i\end{array}\right)\left(\begin{array}{c}j\\i\end{array}\right)}{\left(\begin{array}{c}k+i\\k\end{array}\right)}\bigg].
\end{equation}

It is now easy to show that each coefficient of $(-\frac{1}{N})^i$ is
$0$.  Therefore Eq.~(\ref{A12}) reduces to $0=0$, which is equivalent to the statement that both sides of equation (\ref{numberbasiskA}) are equal.  Therefore we have shown that equation (\ref{generalCA}) is a solution to equation (\ref{numberbasiskA}).  Since equation (\ref{generalCA}) has an undetermined constant for each initial condition $C_{n,n+k}[0]$, this proves by the Uniqueness and Existence theorem that we have found the complete general solution to equation (\ref{numberbasiskA}).
\end{document}